\documentclass[journal]{IEEEtran}

\usepackage{amsfonts}
\usepackage{amsmath}
\usepackage{amssymb}
\usepackage{url}
\usepackage{xcolor}
\usepackage{mathtools}
\usepackage{graphicx}
\usepackage{booktabs}
\usepackage{caption}
\usepackage{subcaption}
\usepackage{cite}
\usepackage{breqn}
\usepackage{mathrsfs}
\usepackage{acronym}
\usepackage{bm}

\graphicspath{{./}}
\newcommand{\eq}[1]{Eq.~\eqref{#1}}
\newcommand{\fig}[1]{Fig.~\ref{#1}}
\newcommand{\tab}[1]{Tab.~\ref{#1}}
\newcommand{\secref}[1]{Section~\ref{#1}}

\newcommand{\mytexttilde}{{\raise.17ex\hbox{$\scriptstyle\mathtt{\sim}$}}}

\begin{document}
\title{Multi-Person Continuous Tracking and Identification from mm-Wave micro-Doppler Signatures}

\author{Jacopo~Pegoraro,~\IEEEmembership{Graduate Student Member,~IEEE,}
        Francesca~Meneghello,~\IEEEmembership{Graduate Student Member,~IEEE,}
        and~Michele~Rossi,~\IEEEmembership{Senior~Member,~IEEE}% <-this % stops a space
\thanks{This work has been supported, in part, by MIUR (Italian Ministry of Education, University and Research) through the initiative "Departments of Excellence" (Law 232/2016) and by the EU MSCA ITN project MINTS ``MIllimeter-wave NeTworking and Sensing for Beyond 5G'' (grant no. 861222).

The authors are with the Department of Information Engineering at the University of Padova, via Gradenigo 36/b, 35131, Padova, Italy. 

Digital Object Identifier 10.1109/TGRS.2020.3019915}}% <-this % stops a space

\markboth{IEEE Transactions on Geoscience and Remote Sensing}%
{Shell \MakeLowercase{\textit{et al.}}: Bare Demo of IEEEtran.cls for Journals}
\maketitle
%\IEEEpubid{0000--0000/00\$00.00 \copyright\ 2013 IEEE}

\IEEEpubid{\begin{minipage}{\textwidth}\ \\[12pt] \centering
		\copyright 2020 IEEE. Personal use is permitted, but republication/redistribution requires IEEE permission.\\
		See http://www.ieee.org/publications\_standards/publications/rights/index.html for more information.
\end{minipage}}

\IEEEpubidadjcol
\begin{abstract}
In this work, we investigate the use of backscattered \mbox{mm-wave} radio signals for the joint tracking and recognition of identities of humans as they move within indoor environments. We build a system that effectively works with multiple persons concurrently sharing and freely moving within the same indoor space. This leads to a complicated setting, which requires one to deal with the randomness and complexity of the resulting (composite) backscattered signal. The proposed system combines several processing steps: at first, the signal is filtered to remove artifacts, reflections and random noise that do not originate from humans. Hence, a \mbox{density-based} classification algorithm is executed to separate the Doppler signatures of different users. The final blocks are trajectory tracking and user identification, respectively based on Kalman filters and deep neural networks. Our results demonstrate that the integration of the \mbox{last-mentioned} processing stages is critical towards achieving robustness and accuracy in \mbox{multi-user} settings. Our technique is tested both on a \mbox{single-target} public dataset, for which it outperforms \mbox{state-of-the-art} methods, and on our own measurements, obtained with a $77$~GHz radar on multiple subjects simultaneously moving in two different indoor environments. The system works in an online fashion, permitting the continuous identification of multiple subjects with accuracies up to $98$\%, e.g., with four subjects sharing the same physical space, and with a small accuracy reduction when tested with unseen data from a challenging \mbox{real-life} scenario that was not part of the model learning phase.
\end{abstract}


\IEEEpeerreviewmaketitle

\section{Introduction} \label{sec:introduction}

\IEEEPARstart{R}{adar} devices for indoor spaces have recently gathered considerable attention. They work by emitting radio waves and analyzing the signal that is reflected by the environment and collected at their receiving antennas. 
In contrast with camera surveillance systems, they are insensitive to poor light conditions and are more privacy preserving, as no video of the scene is collected~\cite{vandersmissen2018indoor}. Radars are also energy efficient compared to other technologies such as LIDARs~\cite{shah2019rf}. 
In this work, we propose a \mbox{multi-person} online identification framework that is based on the analysis of the (reflected) signal received by a \mbox{millimeter-wave} (\mbox{mm-wave}) low power \mbox{frequency-modulated} \mbox{continuous-wave} (FMCW) radar. Our work stems from the observation that reflected signals collected as a subject walks in near proximity of the radar are \mbox{person-specific}, as radio reflections depend on the body shape and, in time, on the movement. As such, they can be used to recognize the identity of humans moving in proximity of the radar device. Our system achieves accuracies as high as $98$\% with four persons moving within a relatively small indoor place.   
Such performance is achieved in an online fashion (continuous tracking and identification), allowing one to recognize user identities  as these share the same physical space, without relying on any visual representation of the scene. We stress that previous work \cite{vandersmissen2018indoor, polfliet2018structured, zhao2019mid} has coped with a \mbox{single-person} identification problem and the \mbox{multi-user} case has only been addressed in an offline fashion through the superposition of multiple \mbox{single-person} signals. In contrast, we build a system that effectively works when multiple persons {\it concurrently} share and {\it freely move} within the same indoor space, directly working on the {\it composite} reflected signal that they generate. 
\IEEEpubidadjcol
To distinguish different persons from their way of walking (gait), we analyze their \emph{micro-Doppler signature} ($\mu$D), i.e., the small scale Doppler effect caused by the human motion.
In the interest of developing a low-complexity system, we first extract $\mu$D features performing range-Doppler (RD) processing (i.e., distance and velocity) of the signal gathered from a {\it single} receiving antenna. 
After that, we address the limitations of RD processing by tackling the so called \mbox{range-Doppler-azimuth} (RDA) space, through the integration of the \mbox{angle-of-arrival} (AoA) of the received radio reflections, estimated using multiple receiving antennas. The AoA information allows resolving targets which are at the same distance from the radar device, and that move with the same velocity; these targets would hardly be separable in the simpler RD space.

The simultaneous identification of multiple targets requires to track and separate the subjects (namely, their contributions to the composite backscattered signal) in order to extract their $\mu$D (temporal) traces. Our technique operates in either the RD or RDA spaces, integrating tracking and identification through the following steps: 1)~\textbf{detection:} random noise is removed and a   \mbox{density-based} clustering algorithm (on either RD or RDA maps) is applied for target detection, 2)~\textbf{tracking:} a dedicated Kalman filtering (KF) algorithm is utilized to track the detected target points in the RD (RDA) space, and 3)~\textbf{identification:} a deep convolutional neural network (DCNN) is exploited to carry out the final identification. We stress that the {\it joint} estimation of user movement (the tracking step 2) and computation of identification features (step 3) is key to correctly disentangle the RD/RDA signals from multiple subjects. As we experimentally verify in \secref{sec:results}, tracking errors and consequent wrong identifications critically depend on this joint processing.

When processing radar data for identification purposes, the analytical models of the propagation and backscattering phenomena often fail to handle the high randomness of \mbox{mm-wave} reflections and hardware \mbox{non-idealities}. To cope with this, we exploit a deep learning architecture (i.e., the DCNN), as it enables a \mbox{data-driven} system training. This technique has become dominant for this type of processing tasks~\cite{vandersmissen2018indoor,seyfiouglu2018deep}.

Differently from previous research efforts, the proposed framework is evaluated by measuring its {\it online} accuracy in the {\it simultaneous} identification of multiple targets, taking into account the additional disturbances, blockages and spurious reflections that are due to the presence of other people, and using experiments designed to reproduce a \mbox{worst-case} scenario for target tracking. To this end, we have emulated a \mbox{real-life} setting, letting subjects walk freely within the scene, at a distance that ranges from $0$ to $18$~meters. In addition, we test the generality of the proposed approach in a room for which no training data was recorded, i.e., this environment is {\it unseen} from the perspective of model learning.

The main contributions of the paper are summarized next.
\begin{enumerate}
	\item We propose a system for the simultaneous indoor identification of multiple targets from $\mu$D signatures of gait using only RD information, reaching an average online accuracy of $95$\% when three subjects walk concurrently within the same physical environment. The approach that we devise for this scenario (RD signal space) works up to long distances ($18$~m) in indoor environments. To the best of our knowledge, no other study in the literature proposes a working system for the considered \mbox{multi-target} online identification task.
	\item We introduce a novel DCNN for $\mu$D processing and quantify its performance improvement with respect to other models presented in the literature by evaluating it on a publicly available dataset (IDRad~\cite{vandersmissen2018indoor}) obtaining an accuracy of $90.69$\%.
	\item We design a new approach for tracking that is robust to trajectory tracking errors thanks to the feedback on the subject identity provided by the DCNN classifier. Our design entails the integration of tracking and identification blocks, which leads to a significant improvement in terms of online identification accuracy.
	\item We show how the proposed processing pipeline can also be applied to RDA data, solving some limitations of the RD signals. This allows one to achieve higher target detection capabilities at the cost of a higher computational complexity and of a reduced detection range. With RDA information, we reach an online accuracy of up to $98$\% for four subjects. The \mbox{RDA-based} system is also evaluated in an {\it unseen} environment, with furniture and static objects, achieving an accuracy of $96$\% on two subjects. To the best of our knowledge, it is the first time that this evaluation is conducted for the mm-wave radar multi-target tracking and identification problem.
\end{enumerate}

The rest of the paper is organized as follows. In \secref{sec:related_work}, the existing literature is reviewed, underlining the novel aspects underpinning our approach. In \secref{sec:dmaps}, the FMCW radar signal model and the computation of RD, RDA maps and $\mu$D signatures is detailed. The new framework is thoroughly presented in \secref{sec:proposed_algorithm}. In \secref{sec:results}, experimental results are presented, while concluding remarks are given in \secref{sec:conclusions}. 

\section{Related work}
\label{sec:related_work}

Human identification from radar sensors is a research theme that is rapidly gaining momentum. Some papers target the classification of the subject identity from the $\mu$D signature of gait using radio signals~\cite{vandersmissen2018indoor, cao2018radar,yang2019person,abdulatif2019person,chen2018personnel,jalalvand2019radar,polfliet2018structured}. 
%Other studies focus on human activity recognition or vital signs detection \cite{luo2019human,trommel2016multi,seyfiouglu2018deep,perez2018single,weishaupt2018vital,hsieh2015uwb,kim20191d}.
%The general idea is the recognition of different human movements and activities 
Other studies focus on human activity recognition from the backscattered radio signal for security or \mbox{smart-home} applications~\cite{luo2019human,trommel2016multi,seyfiouglu2018deep}. Respiration rate and heartbeat can also be tracked, as they cause a detectable movement of the subject's chest~\cite{weishaupt2018vital,hsieh2015uwb}. As the focus of this paper is on gait recognition and person identification, in the following we briefly review the most important contributions on this topic. 

In~\cite{cao2018radar}, the authors employ for the first time a classifier based on the deep CNN AlexNet~\cite{krizhevsky2012imagenet} to identify a person from her/his $\mu$D signature of gait, reaching an accuracy of about $97$\% with four subjects. Differently from our setup, their experiments take place in an outdoor environment, where correlated noisy reflections from static objects are typically weak: walls in indoor environments are significantly close to the target of interest in most scenarios, and they cause the noise level to increase making the extraction of the useful signal features much harder.

Chen \emph{et al.} \cite{chen2018personnel} utilize a \mbox{multi-static} radar with three nodes and a \mbox{pre-trained} deep CNN for image recognition, in order to detect whether a person carries a weapon or to identify a person between two subjects. The authors of~\cite{yang2019person} address identification using the $\mu$D signature of six different movements including walking and running. Running turned out to be the most discriminative action, providing an identification accuracy of $95.21$\% with $15$ subjects. In~\cite{abdulatif2019person}, instead, a treadmill placed at different distances from the radar device is used, and a ResNet50~\cite{he2016deep} neural network is trained to classify $22$ subjects.

The above studies focused on simplified experimental scenarios, where the person was required to walk on a straight line, in a radial direction from the radar device. This approach can be useful to simplify the classification task, by making gait features more evident, but it is not realistic and lacks the generality that would be required by a practical application. In our current work, we focus on a more realistic setup, letting the subjects walk in an unconstrained, free manner within the monitored physical space.

Vandersmissen \emph{et al.}~\cite{vandersmissen2018indoor} train a CNN classifier on a dataset featuring five subjects who randomly walk in two different rooms, in an attempt to implement a more robust learning phase. However, each subject needs to be alone in the room in order for the system to work, as no method to separate the different target contributions in the backscattered signal is provided. This heavily limits the applicability of the proposed algorithm to real situations, where multiple targets are likely to share and concurrently move in the physical space. The same authors also propose two improvements over their algorithm, to improve its accuracy, but the \mbox{single-target} limitation is still present~\cite{jalalvand2019radar, polfliet2018structured}.

A first attempt at performing \mbox{multi-subject} identification can be found in~\cite{zhao2019mid}, where \mbox{$3$-dimensional} radar point clouds obtained by RDA processing are used in place of $\mu$D signatures, in combination with a recurrent neural network with long short-term memory (LSTM) cells for \mbox{a-posteriori} identification. The overall accuracy obtained for $12$ subjects is around $89$\%, and evidence that the system is able to distinguish between two subjects is provided. However, no evaluation of the accuracy is conducted when more than $2$ subjects share the same environment, and no evaluation is conducted to assess the generality of the classifier by testing it in a different indoor environment (e.g., a new room) after the training.
The sparsity of radar \mbox{point-cloud} data can become a source of inaccuracy when a high number of subjects has to be tracked, due to failures in the clustering procedure. To date, no method exists to deal with the superposition of the signal clusters caused by the proximity of the subjects, thus limiting the working range of identification systems to a radius of \mbox{$3-5$~meters}. 

In this work, we improve over previous studies by first showing the feasibility of identifying multiple persons only using RD information, with a lightweight processing workflow and limited hardware requirements. Further, we extend the proposed system to deal with RDA data, in case a higher detection performance is required, e.g., to handle more targets, or in case precise tracking of the subjects in the $x - y$ space is sought. We also show how the complex task of reliably separating the different user's reflections (especially in RD images) can be successfully tackled by feeding back the identification output into the user's \mbox{trajectory tracking} module, combining these two processing stages. Improvements and drawbacks of our approach are duly quantified and discussed.

\section{mm-wave radar signal model}
\label{sec:dmaps}

\begin{figure*}[t!]
	\begin{center}   
%		\textbf{Your title}\par\medskip
		\centering
		\subcaptionbox{RD map\label{fig:radar-maps-rd}}[3cm]{\includegraphics[height=4.5cm]{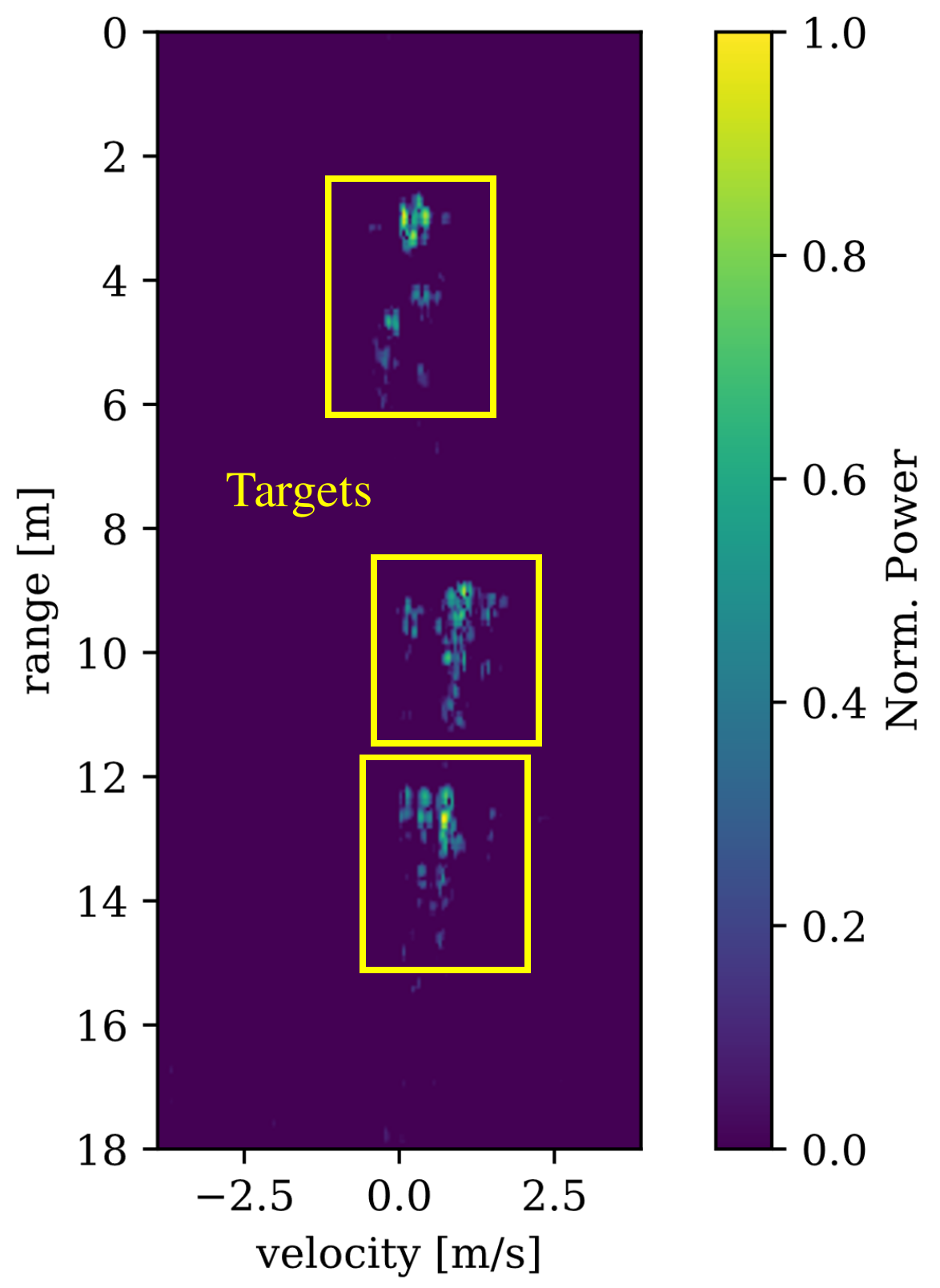}}
		\subcaptionbox{RDA map\label{fig:radar-maps-rda}}[7cm]{\includegraphics[width=6cm]{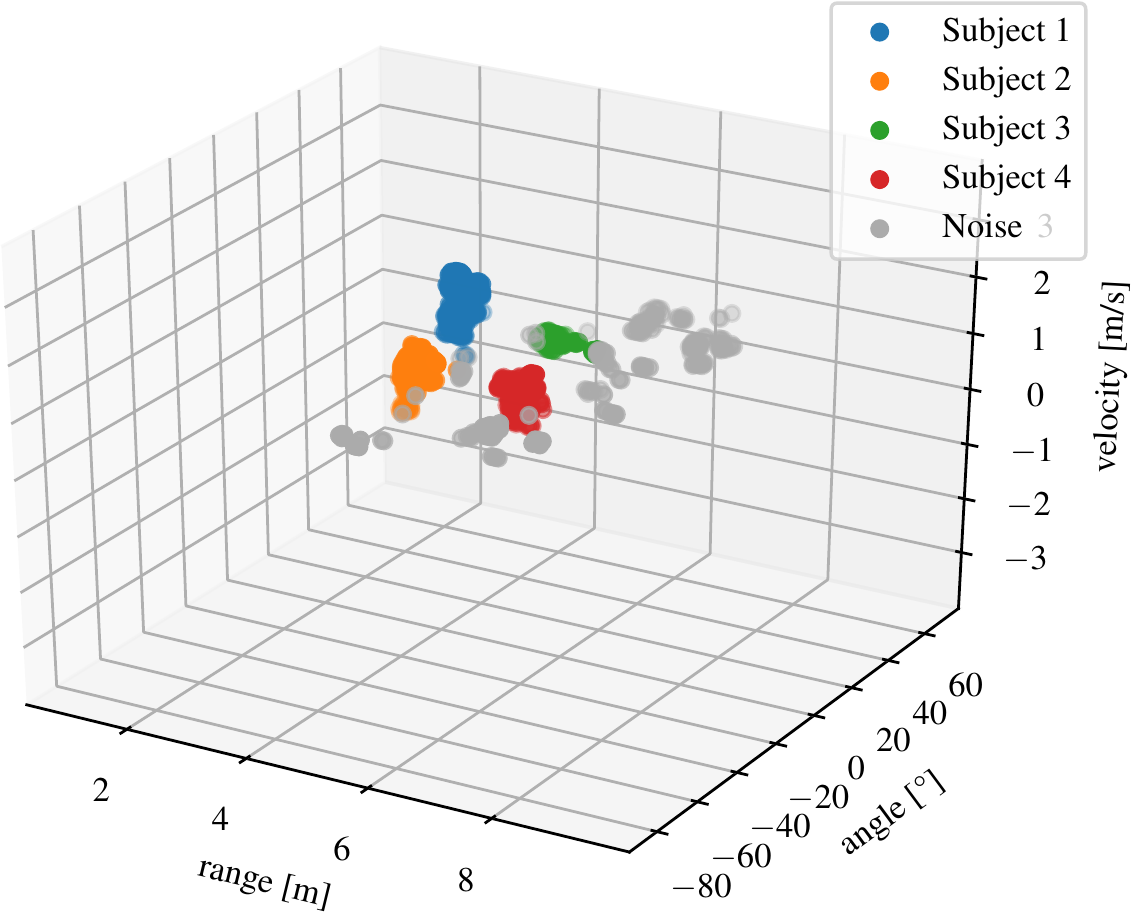}}
		\subcaptionbox{$\mu$D signature\label{fig:radar-maps-mud}}[7cm]{\includegraphics[width=7cm]{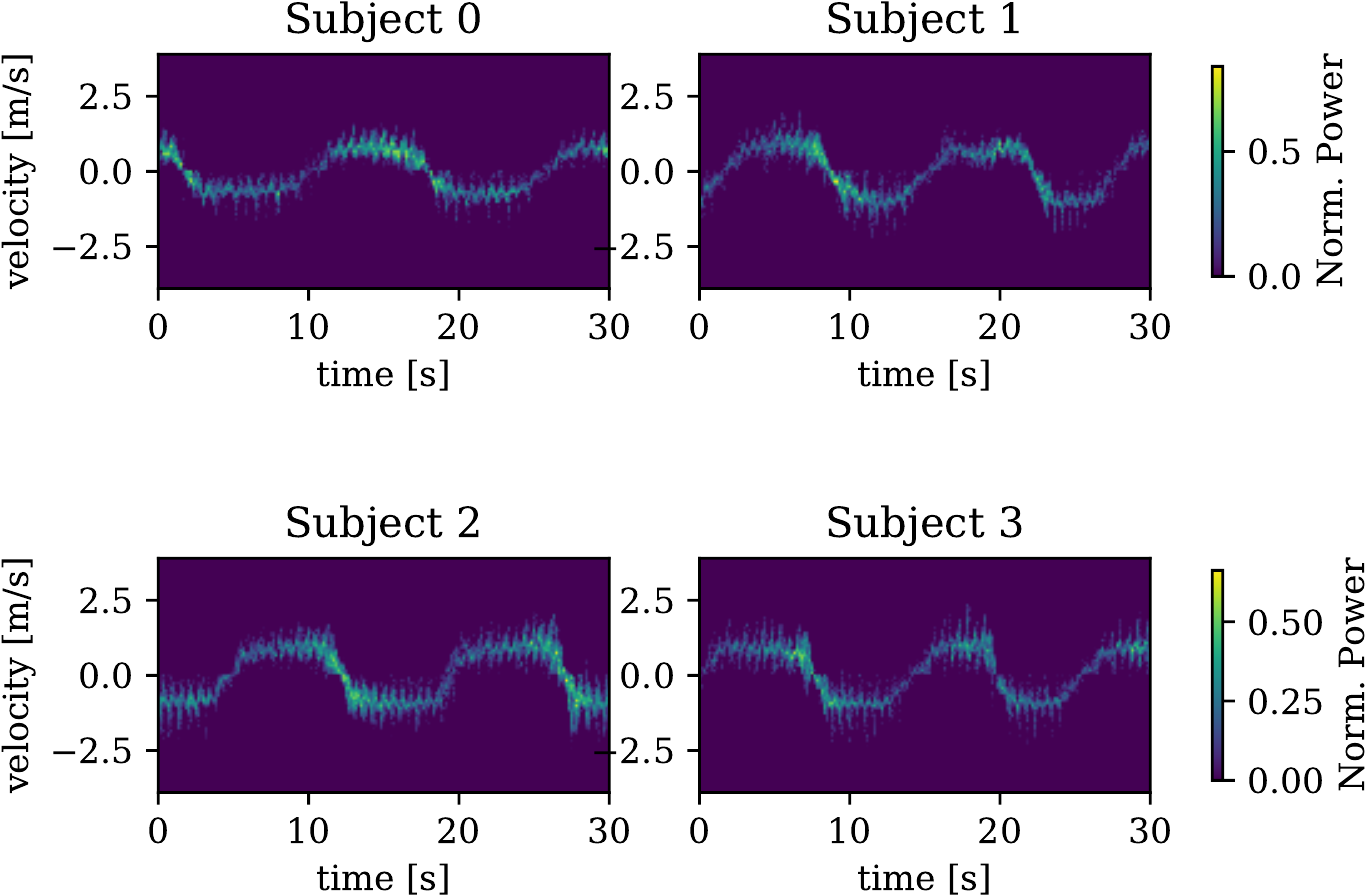}}
		\caption{Visual representation of the RD, RDA maps and $\mu$D signature after a thresholding operation is applied. In the RD map $3$ targets are present, while in RDA and $\mu$D $4$ targets are considered.}
		\label{fig:radar-maps}
	\end{center}
\end{figure*}

A FMCW radar allows the joint estimation of the distance and the radial velocity of the target with respect to the radar device.
This is achieved by transmitting sequences of \emph{chirps}, i.e., sinusoidal waves with frequency that varies in time, and measuring the frequency shift of the backscattered signal at the receiver.

In this paper, we use a linear FMCW (LFMCW) radar for which the frequency of the transmitted chirp signal (TX) is linearly increased from a base value $f_o$ to a maximum $f_1$ in $T$ seconds. Defining the bandwidth of the chirp as \mbox{$B = f_1 - f_o$}, bandwidth $B$ and transmission duration $T$ are related through \mbox{$\zeta = B/T$}, and the transmitted signal can be expressed as  
%In the following we assume that the transmitted (TX) chirp signals have duration of $T$ seconds and are repeated every $T_{\rm rep}$ seconds; the frequency is linearly increased from a base value $f_o$ to a maximum $f_1$. We denote by \mbox{$B = f_1 - f_o$} the bandwidth of the chirp pulse and \mbox{$\zeta = B/T$} the slope of the dependency between time and frequency.
%The expression of a transmitted chirp signal is
\begin{equation}
s(t) = \exp \left\{ j 2\pi \left(f_o + \frac{1}{2}\zeta t \right)t \right\}, \, 0 \leq t \leq T.
\end{equation}
The chirps are transmitted every $T_{\rm rep}$ seconds in sequences of $P$ chirps each, so that the total duration of a transmitted sequence is $PT_{\rm rep}$.
%The chirps are transmitted every $T_{\rm rep}$ seconds.
%We call $P$ the number of chirps for each sequence, so the total duration of a transmitted sequence is $PT_{\rm rep}$.
At the receiver, a mixer combines the received signal (RX) with the transmitted one, generating the intermediate frequency (IF) signal, i.e., a sinusoid whose instantaneous frequency is the difference between the frequencies of the TX and RX signals.
%The output of the mixer is a sinusoid whose instantaneous frequency is the difference between the frequencies of the TX and RX signals, called Intermediate Frequency (IF) signal.
Each chirp is sampled with sampling period $T_s$ (referred to as \emph{fast time} sampling) obtaining $N$ points, while $P$ samples, one per chirp from adjacent chirps, are taken with period $T_{\rm rep}$  (\emph{slow time} sampling).

The use of multiple-input multiple-output (MIMO) radar devices allows the additional estimation of the AoA of the reflections, by computing the phase shifts between the receiver antenna elements due to their different positions (i.e., their different distances from the target). This is referred to as \emph{spatial} sampling, and enables the localization of the targets in the physical space using polar and cartesian coordinates.
In the present work we employ a \textit{linear} receiver antenna array, i.e., the RX antennas are aligned along a single dimension and spaced apart by a distance $\delta$.

\subsection{Range, Doppler and azimuth information} \label{sec:rdmap}

The transmitted signal hits the target at some spatial point, generating a backscattered signal that can be detected at the receiver.
This reflected signal is equal to the transmitted waveform with a delay $\tau$ that depends on the distance between the target and the radar, their relative radial velocity, and on the additional distance due to the different positions of the receiving antenna elements. 
Considering the most general case where $Q$ targets are present in the radar illumination range and $L$ antennas are available at the linear receiver array, and indicating with $c$ the speed of light, letting $R_q$, $v_q$ and $\theta_q$ respectively be the range, velocity and azimuth angle with respect to the device of target $q$, the delay measured at antenna element $\ell$ for the signal reflection coming from target $q$ can be computed as

\begin{equation}\label{eq:3d-delay}
\tau_{\ell q} = \frac{2(R_q + v_q t) + \ell\delta\sin \theta_q}{c}.
\end{equation}
%where $c$ is the speed of light, $R_q$, $v_q$ and $\theta_q$ are respectively the range, velocity and azimuth angle with respect to the device of target $q$.
After mixing and sampling, the IF signal is expressed as~\cite{patole2017automotive}
\begin{equation}\label{eq:3d-signal}
y(n, p, \ell) = \sum_{q=0}^{Q-1} \alpha_q \exp\left\{j2\pi \phi_{q}(n, p, \ell)\right\} + w(n, p, \ell),
\end{equation}
where $\alpha_q$ is a coefficient that accounts for the attenuation effects due to the antenna gains, path loss and radar cross section (RCS) of the target and $w$ is a Gaussian noise term. 
The phase $\phi_{q}(n, p, \ell)$ depends on the target, the fast time, slow time and spatial sampling indices.
By neglecting the terms giving a small contribution, its approximate expression can be written by introducing the quantities \mbox{$f_{d_q}=2f_o v_q/c$} and \mbox{$f_{b_q}=2\zeta R_q/c$}, which respectively represent the Doppler frequency and the \emph{beat} frequency of the signal reflected from target $q$,
%Its expression can be written as follows \cite{patole2017automotive}
\begin{equation}\label{eq:3d-phase}
\phi_{q}(n, p, \ell) \approx \frac{2f_oR_q}{c} + f_{d_q} p T_{\rm rep} + \frac{f_o \ell \delta \sin \theta_q}{c} + \left(f_{d_q} + f_{b_q} \right)nT_s.
\end{equation}
Samples of $y$ can be arranged into a \mbox{3-dimensional} tensor called \emph{radar data cube}, that contains all the information provided by the radar device for a given time frame. The frequency shifts of interest, which reveal the target range, velocity and angular position, can be extracted after applying a discrete Fourier transform (DFT) along the fast time, slow time and spatial dimension (beamforming). In the resulting signal, the position of the peak along the fast time dimension reveals the frequency of the IF signal \mbox{$f_{d_q} + f_{b_q} \approx f_{b_q}$}, the peak along slow time gives the Doppler frequency $f_{d_q}$. From the peak of the DFT along the spatial dimension we get the phase shift due to the angular displacement of the target, $\varphi_{a_q}$.
The desired quantities are then estimated as follows (we indicate with the symbol $\Delta$ the corresponding resolution)
\begin{equation}\label{eq:range-est}
\hat{R}_q = \frac{f_{b_q} c}{2 \zeta}, \quad \Delta\hat{R}_q = \frac{c}{2B},
\end{equation}
\begin{equation}\label{eq:vel-est}
\quad \hat{v}_q = \frac{f_{d_q} c}{2f_o}, \quad \Delta\hat{v}_q = \frac{c}{2f_oPT_{\rm rep}},
\end{equation}
\begin{equation}\label{eq:angle-est}
\quad \hat{\theta}_q = \sin^{-1} \left( \frac{\varphi_{a_q} c}{2 \pi \delta f_o} \right), \quad \Delta\hat{\theta}_q = \frac{c}{2\delta L\cos(\hat{\theta}_q)} .
\end{equation}
In the following, the radar cube after applying the DFT in the three dimensions will be referred to as \emph{range-Doppler-azimuth map} (RDA). An example of the RDA map for four subjects is shown in \fig{fig:radar-maps-rda}.

In the case of a single receiving antenna, spatial sampling is not possible, and we can only estimate the range and velocity of the targets with the same approach used above, with the difference that \eq{eq:3d-delay} and \eq{eq:3d-phase} do not depend on the antenna element $\ell$. The result of the \mbox{2-dimensional} DFT is called \emph{range-Doppler map} (RD), see the example in \fig{fig:radar-maps-rd}.

\subsection{$\mu$-Doppler map}

Human targets present different moving parts, therefore their overall motion is more complex than just translation.
The small-scale vibrations or rotations of their body parts introduce a Doppler shift that is time dependent and that can be represented as a frequency modulation on the reflected signal, which carries unique features depending on the specific target considered. A model for this phenomenon is presented in \cite{chen2000analysis, chen2006micro}, where it is shown that the sensitivity to $\mu$D effects is higher when using small wavelenghts: \mbox{mm-wave} radios are therefore more suited for applications where fine grained information is needed.

The extraction of the $\mu$D signature from the received signal can be performed by computing a short-time Fourier transform (STFT) on the \mbox{slow-time} sampled waveform to estimate the power spectral density (PSD) along the Doppler dimension, as done in \cite{abdulatif2019person}.
An alternative is to compute the RD (or RDA) map first, and subsequently integrate along the range and angular (or range only) dimensions  \cite{vandersmissen2018indoor}, as shown in \fig{fig:radar-maps-mud}. This second option is computationally more expensive, but it is preferred here because the RD (respectively RDA) map can be used to locate the targets and separate their contributions in a 2D (resp. 3D) space, while this separation would be very hard from the $\mu$D spectrogram, as it lacks the range (resp. range and angle) information.

\begin{figure*}[t!]
	\begin{center}   
		\includegraphics[width=15cm]{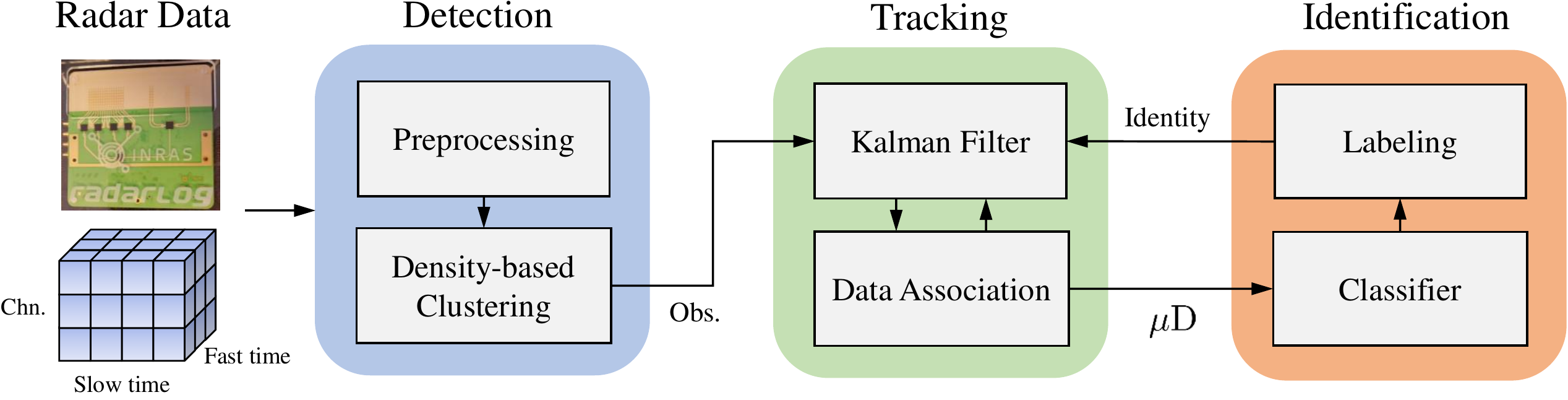} 
		\caption{Signal processing workflow.}
		\label{fig:proc_pip}
	\end{center}
\end{figure*}

\section{Proposed algorithm} 
\label{sec:proposed_algorithm}

In this section, we offer a general overview of the proposed algorithm. The blocks that are presented here are used for both RD and RDA processing, with minor differences in the implementation details of each algorithm, due to the different properties of the two maps.

\subsection{Overview of the signal processing pipeline}

The extraction of the gait features from the $\mu$D spectrogram can be very difficult, and the results are heavily affected by environment and hardware \mbox{non-idealities}. In addition, in the case of multiple targets, the $\mu$D is a composite temporal signal resulting from the superimposed contributions of all moving entities. The separation of such contributions is very hard, whereas it is easier in the RD or RDA spaces as the reflections from different users are further spaced out by the distance of the users from the radar (RD) or by their distance and angle of arrival (RDA), resulting in point clouds as shown in Fig.~\ref{fig:radar-maps}. For this reason, our dynamic processing framework works on either the RD or RDA spaces, through the following steps (see Fig.~\ref{fig:proc_pip}).
\begin{enumerate}
	\item \textbf{Detection}. At first, a \mbox{pre-processing} step is applied to the raw data at the output of the radar mixer, to remove static reflections and noise (see \secref{sec:preprocessing}). Hence, a clustering scheme from the family of ``density-based spatial clustering for applications with noise'' \mbox{(DBSCAN)} algorithms is executed to separate the RD/RDA contributions from distinct subjects from the composite signal (see \secref{sec:hdbscan}).

    \item \textbf{Tracking}. A Kalman filter operating on subsequent RD or RDA frames is applied to obtain a reliable estimation of the true subject's state (i.e., its location, see \secref{sec:kalman}). The association of the RD/RDA clusters detected in the current \mbox{time-frame} with the right user trajectories is performed using the Hungarian algorithm (see \secref{sec:hungarian}).

	\item \textbf{Identification}. Feature extraction and user identification are performed with a DCNN model based on inception blocks (IBs) that takes as input portions of the $\mu$D spectrogram of each subject (obtained from the RD/RDA data of the subject, after the use of DBSCAN and trajectory tracking). In case tracking fails and the RD/RDA clusters of some subjects cannot be  separated, the DCNN output is used to \mbox{re-establish} the correct labeling of the targets, by feeding back the identity information to the trajectory tracking block (see \secref{sec:correction} for details).
\end{enumerate}

\mbox{Multi-person} identification from backscattered \mbox{mm-wave} signals presents several challenges. First, an effective and reliable separation of the different targets is difficult to achieve due to the high level of randomness in \mbox{mm-wave} indoor propagation environments. Second, a robust classification based on $\mu$D signatures requires high generalization capabilities from the DCNN identity classifier. Indeed, we seek to differentiate subjects from their way of moving rather than from properties that may be less \mbox{person-specific}, such as their average walking speed. A distinctive and key feature of the proposed approach is the dynamic integration of trajectory tracking and identification, which allows correcting trajectory tracking errors based on the output of the identification block. As a result, our system is suited to online processing, is robust to the superposition of user clusters in the RD/RDA spaces, to variable walking speeds, to fake targets due to reflecting objects/surfaces, to classification instability and to targets appearing on (disappearing from) the scene. 

\subsection{Notation}
The system operates at discrete time increments, $t=1,2, \dots T$, where time steps have a fixed duration of $\Delta t$ seconds,  corresponding to the radar frame period. In the remainder, the sequential evolution of the algorithms is interchangeably expressed in terms of time steps and radar frames. The RD/RDA clusters detected in the current time step $t$ are marked with indices $d = 0, 1, \dots, D_t -1$ and are $D_t$ in total. Similarly, the $K_t$ trajectories that are currently maintained by the trajectory tracking block are indexed using variable $k = 0, 1, \dots, K_t -1$. With $U$, we indicate the number of classes (identities) on which the system is trained, i.e., the identities that will be recognized as known, represented through index $u = 0,1,\dots, U$. Boldface, capital letters refer to matrices, e.g., $\boldsymbol{X}$, with elements $X_{ij}$, whereas boldface lowercase letters refer to vectors, e.g., $\boldsymbol{x}$. Symbol $\otimes$ denotes the Kronecker product between matrices, $\boldsymbol{X}^{-1}$ denotes the inverse of matrix $\boldsymbol{X}$, and $\boldsymbol{x}^T$ denotes the transpose of vector $\boldsymbol{x}$. $\mathcal{N}(\mu, \sigma^2)$ indicates a Gaussian random variable with mean $\mu$ and variance $\sigma^2$. 

\subsection{Pre-processing} \label{sec:preprocessing}
The pre-processing involves two different phases, namely removal of static reflections and denoising.
\subsubsection{Removal of Static Reflections}
This is the first block in the processing pipeline: it receives as input the raw radar data, i.e., the radar cube containing the $3$-dimensional signal (see \eq{eq:3d-signal}) that the radar outputs at every time step. As discussed in Section \ref{sec:rdmap}, DFT is applied to this signal to obtain the \mbox{RD} or \mbox{RDA} map. In the RD case, only one channel is collected (one receiving antenna), the DFT is applied along the range dimension first and then along the Doppler dimension, resulting in a matrix containing range and Doppler information on the targets. In the RDA processing case, an additional DFT along the angular dimension is computed. Before the DFT, a Hanning window is applied along each dimension. The RD and RDA maps are further processed to remove reflections from static targets. As fixed objects are mapped into a vertical line in correspondence of the $0$~m/s velocity value, we remove their contributions by cutting the Doppler channels related to negligible velocities from both the RD and RDA maps.
This processing step is of key importance, as the static clutter would dominate the RDA maps if not removed, causing the reflections from the subjects to be merged with static contributions with consequent severe difficulties in the tracking process. As an additional benefit, the algorithm becomes less dependent on the environment characteristics.

\subsubsection{Denoising}

Denoising is applied in two phases. In the first phase, a received power threshold is applied along the range dimension, keeping only the signal values that lie above it. The threshold is decreased linearly in the logarithmic domain as the range increases, going from $-97$~dBm at minimum range to $-107$~dBm at maximum range. This is motivated by the fact that targets further away from the radar device would be penalized by using a fixed threshold due to the smaller power they receive. In case of RDA processing, a further thresholding is applied along the angular dimension, discarding the angular bins where the received power level is weaker than $15$~dB with respect to the peak value. This is implemented to mitigate the effects of the side lobes generated by the beamforming procedure. The resulting data points represent the locations in the $2$-dimensional (RD) or $3$-dimensional (RDA) maps, where a sufficiently high reflected power is received. These points represent candidate reflections from the targets.

\subsection{Target clustering in RD/RDA spaces -- DBSCAN} \label{sec:hdbscan}

\mbox{Density-based} clustering, as opposed to \emph{distance}-based clustering, groups input samples depending on their density. 
One of the most widely used algorithms belonging to this category is \mbox{DBSCAN}~\cite{ester1996density}, which has been previously applied to clusterize radar point clouds in~\cite{wagner2017radar, zhao2019mid}. The algorithm operates a sequential scanning of all the data points, expanding a cluster until a certain density condition is no longer satisfied. It requires one to specify two input parameters, $\epsilon$ and $m_{\rm pts}$, respectively representing a radius around each point and the minimum number of other points inside of it to satisfy the density condition. In this work, we use $\epsilon=0.04$ and $m_{\rm pts}=40$. Each point of the radar map, after denoising, is mapped onto a vector of coordinates \mbox{$\boldsymbol{p}_i = \left[r_i, v_i \right]^T$} (range and velocity) for RD processing and \mbox{$\boldsymbol{p}_i = \left[r_i, v_i, \theta_i \right]^T$} (range, velocity and angle) for RDA processing, with an associated received power $P_{\rm RX}(\boldsymbol{p}_i)$.
To simplify the selection of the distance threshold parameter, $\epsilon$, the range, angle and velocity coordinates of the points $\boldsymbol{p}_i $ are normalized in the interval $[0,1]$ before the actual clustering step.
\mbox{DBSCAN} is applied on the normalized set of points: some, having low density, are classified as noise and discarded, while a partition of the remaining ones is outputted at each time step $t$. We denote by $C_0, C_1, \dots, C_{D_t-1}$ the resulting clusters, one for each of the $D_t$ detections. After the clustering operation, the point clouds are \mbox{re-mapped} onto the original range of values. For each cluster, we select its centroid as a noisy observation of the true coordinates (range and velocity for RD, range, velocity and angle for RDA) of the person. Centroids $\boldsymbol{z}_d, ~d=0,1,\dots, D_t-1$, are computed by weighting each cluster point by the corresponding normalized reflected power value, namely,
\begin{equation}
\boldsymbol{z}_d = \frac{\sum_{\boldsymbol{p}_i \in C_d}\boldsymbol{p}_i P_{\rm RX}(\boldsymbol{p}_i)}{\sum_{\boldsymbol{p}_j \in C_d}P_{\rm RX}(\boldsymbol{p}_j)}.
\end{equation}
In this way, the centroid tends towards those points with a higher power, assigning them more importance in representing the actual target position. Note that, DBSCAN clustering performs the detection of the clusters by solely operating on the present time step, i.e., points in previous time steps are not considered. While this is desirable, as it leads to a {\it low complexity} clustering algorithm, it presents some drawbacks. In fact, not all the clusters that are detected in any specific time step may represent actual subjects, but noisy reflections and ghost targets often appear (at random) in the RD/RDA space. When their power is comparable with that of the actual target reflections, DBSCAN may enroll them among the detected clusters. To compensate for this, we use a further tracking procedure, described in the following \secref{sec:kalman}, that analyzes the movement of the clusters in the RD/RDA space across subsequent frames. This allows detecting and removing spurious clusters, as these are likely to appear (and disappear soon after) at random times, whereas the clusters generated by actual targets tend to be persistent across frames. 

\subsection{Trajectory tracking -- Kalman filter} \label{sec:kalman}

Trajectory tracking is carried out by applying a discrete Kalman filter (KF) on the past positions of the targets, which are represented by the cluster centroids \mbox{$\boldsymbol{z}_0, \dots, \boldsymbol{z}_{D_t -1}$}. Note that the number of maintained trajectories at the \textit{beginning} of time step $t$, $K_{t-1}$, may differ from the number of clusters $D_t$ detected by \mbox{DBSCAN}, due to errors in the clustering procedure or to subjects entering or leaving the monitored environment. These facts need to be carefully handled through dedicated strategies, which are detailed in \secref{sec:hungarian}. Next, the KF tracking procedure is presented for a single trajectory, but this same procedure is applied in parallel to each trajectory. Also, for improved clarity, the RD and RDA processing cases are  treated separately.
\subsubsection{RD system model}
The KF model relates the actual distance (from the radar device) and velocity of the target, $\boldsymbol{x}_t = \left[ r_t, v_t \right]^T$, i.e., the hidden system state, to the centroid values $\boldsymbol{z}_t$, i.e., the (noisy) observations. The model of motion is defined by two matrices, $\boldsymbol{F}$ and $\boldsymbol{H}$. $\boldsymbol{F}$ is the transition matrix, relating the system state in the current time step $\boldsymbol{x}_t$ to $\boldsymbol{x}_{t-1}$, while $\boldsymbol{H}$ is the observation matrix, which relates $\boldsymbol{z}_t$ to $\boldsymbol{x}_t$. Referring to $\boldsymbol{u}_t$ and $\boldsymbol{r}_t$ as the process noise and observation noise, respectively, a dynamic model of the system is obtained as follows
\begin{equation}\label{eq:stateup}
\boldsymbol{x}_{t} = \boldsymbol{F}\boldsymbol{x}_{t-1} + \boldsymbol{u}_t,
\end{equation}
\begin{equation}\label{eq:obs-state}
\boldsymbol{z}_{t} = \boldsymbol{H}\boldsymbol{x}_{t} + \boldsymbol{r}_t,
\end{equation}
Assuming a constant velocity model, the transition and observation matrices are
\begin{equation}
\boldsymbol{F} = 
\left[
\begin{array}{cc}
1 & \Delta t \\ 0 & 1
\end{array}
\right],
\end{equation}
\begin{equation}
\boldsymbol{H} = \boldsymbol{I}_2,
\end{equation}
where $\boldsymbol{I}_2$ is a $2\times 2$ identity matrix. We assume the process noise $\boldsymbol{u}_{t} $ to be caused by a random acceleration $a_t$ that follows a Gaussian distribution with $0$ mean and variance $\sigma_a^2$, i.e., $a_{t} \sim \mathcal{N}(0, \sigma_a^2)$, leading to
\begin{equation}\label{eq:proc_noise}
\boldsymbol{u}_{t} = \boldsymbol{g}a_{t},
\end{equation}
\begin{equation}\label{eq:matrix_G}
\boldsymbol{g} = 
\left[
\begin{array}{c}
\frac{1}{2} \Delta t^2\\  \Delta t
\end{array}
\right].
\end{equation}
The process noise covariance matrix is computed as
\begin{equation}\label{eq:proc-cov}
\boldsymbol{Q} = E\left[ \boldsymbol{u}_t \boldsymbol{u}_t^T \right] = \sigma_a^2\boldsymbol{g}\boldsymbol{g}^T,
\end{equation}
while the observation noise covariance matrix is
\begin{equation}\label{eq:obs-cov}
\boldsymbol{R} =  E\left[ \boldsymbol{r}_t\boldsymbol{r}_t^T\right] = 
\left[
\begin{array}{cc}
\sigma_r^2 & 0\\ 0 & \sigma_v^2
\end{array}
\right].
\end{equation}
Suitable values for $\sigma_a, \sigma_r$ and $\sigma_v$ are difficult to compute analytically. In this work, we determined them empirically from experimental observations, obtaining $\sigma_a = 0.6$~m/s$^2$, $\sigma_r=0.1$~m and $\sigma_v = 0.5$~m/s.

A new KF model is initialized for each detected cluster in the first frame received by the radar. In successive frames, the trajectories are maintained through the KF \mbox{predict-update} steps, computing the estimates of the state $\hat{\boldsymbol{x}}_t$ and state covariance matrix $\hat{\boldsymbol{P}}_t$, from which the estimated posterior distribution of the state is derived as $\hat{p}(\boldsymbol{x}_t | \boldsymbol{z}_{1}, \dots, \boldsymbol{z}_{t}) = \mathcal{N}(\hat{\boldsymbol{x}}_t, \hat{\boldsymbol{P}}_t)$ \cite{kalman1960new}.

\subsubsection{RDA system model}
In the RDA case, tracking is only performed using the observations on range and azimuth, as the introduction of radial velocity in the model would cause the system to become too \mbox{non-linear} to obtain reliable estimates using KF. In detail, the observation vector $\boldsymbol{z}_t$ contains the range and the angular position of the target, $\boldsymbol{z}_t = \left[ r_t, \theta_t \right]^T$. The system state is defined as $\boldsymbol{x}_t = \left[ x, v_x, y, v_y \right]^T$, where $x$ and $y$ are the target cartesian coordinates, and $v_x$ and $v_y$ the velocities along the two axes.  The resulting \mbox{non-linear} model is 
\begin{equation}\label{eq:stateup-nonl}
\boldsymbol{x}_{ t} = \boldsymbol{F}\boldsymbol{x}_{ t-1} + \boldsymbol{u}_t,
\end{equation}
\begin{equation}\label{eq:stateup-nonl-2}
\boldsymbol{z}_{ t} = h\left(\boldsymbol{x}_{ t}\right) + \boldsymbol{r}_t,
\end{equation}
with \mbox{$h\left(\boldsymbol{x}_{ t}\right) = \left[ \sqrt{x^2 + y^2}, ~\arctan \left(y/x\right)\right]^T$}. To handle the \mbox{non-linearity} in \eq{eq:stateup-nonl-2}, upon receiving a new observation $\boldsymbol{z}_{t}$, we compute a transformed observation vector \mbox{$\boldsymbol{z}^{\prime}_{t}=\left[r_t \cos \theta_t,~ r_t \sin \theta_t \right]^T$}. Using $\boldsymbol{z}^{\prime}$, the model becomes linear as in \eq{eq:stateup}, \eq{eq:obs-state}, with matrices
\begin{equation}
\boldsymbol{F} =  \boldsymbol{I}_2 \otimes 
\left[ 
\begin{array}{cc}
1 & \Delta t \\ 0 & 1
\end{array}
\right],
\end{equation}
\begin{equation}
\boldsymbol{H} =
\left[ 
\begin{array}{cccc}
1 & 0 & 0 & 0 \\ 
0 & 0 & 1 & 0 
\end{array}
\right].
\end{equation}
The covariance matrices of the process and observation noises are
\begin{equation}
\boldsymbol{Q} =  \boldsymbol{I}_2 \otimes \sigma_a^2\boldsymbol{g}\boldsymbol{g}^T,
\end{equation}
\begin{equation}
\boldsymbol{R} = 
\left[
\begin{array}{cc}
\sigma_x^2 & 0\\ 0 & \sigma_y^2
\end{array}
\right].
\end{equation}
Again, a direct computation of the noise variances is difficult to obtain, so we used the empirical values for human subjects proposed in \cite{wagner2017radar}: $\sigma_a = 8$~m/s$^2$, $\sigma_x = \sigma_y = 0.5$~m/s$^2$. The linear equations of the predictions and update steps are the same as in the linear KF from the case of RD processing, thanks to the use of the transformation (polar coordinates).

The constant velocity model we used has provided good approximations of the movement of a human walking target: with movements speeds in the order of $1$~m/s and a frame rate of $15$~fps, the KF was able to track the targets reliably.

\subsection{Matching trajectories to clusters -- Hungarian algorithm}\label{sec:hungarian}
  
To match trajectories to clusters, we use an approach based on the nearest neighbor standard filter (NNSF). At each frame, we must associate the $D_t$ new cluster detections with the $K_{t-1}$ previous trajectories, which is a combinatorial problem. The procedure consists in two steps, first we compute a \mbox{$K_{t-1} \times D_t$} cost matrix $\boldsymbol{J}$ that relates trajectories at time step $(t-1)$ with cluster detections at time step $t$. Each element of $\boldsymbol{J}$, $J_{ij}$, encodes the cost of associating trajectory $i$ with cluster detection $j$. Given the slightly different properties of RD and RDA data, we found that the best choice for the cost function differs in the two cases, as described below.

\subsubsection{RD cost matrix}
in the RD case, from each target state $\boldsymbol{x}_i$ we define a box $B_i$ to contain the subject reflections, centered on the state and with fixed dimensions $h_{B}$ and $w_{B}$. We assume that, given the high frame rate with respect to the velocity of the subjects, over two subsequent frames the box with reflections from a given target should significantly overlap with her/his box at the previous time step.
Let $B_i$ and $B_j$ respectively represent the box of the cluster that was associated with trajectory $i$ at the previous time step $(t-1)$ and the one associated with a new target detection $j$ at the current time step $t$, centered on $\boldsymbol{z}_j$. The cost of the association between trajectory $i$ and the newly detected cluster $j$ is computed via the negative intersection over union (IOU) score, defined as
\begin{equation}
J^{\rm RD}_{ij} = -\mbox{IOU}(B_i, B_j) = - \frac{\mbox{Area}(B_i \cap B_j)}{\mbox{Area}(B_i \cup B_j)}.
\end{equation}
The idea underpinning this, is that the more the two boxes overlap, the more likely they will be representing two clusters containing the reflected signal components from the \textit{same} target user as she/he moves from $(t-1)$ to $t$. 

\subsubsection{RDA cost matrix} 
in the RDA case, the cost matrix elements are defined as the squared Mahalanobis distance between the predicted observation from the trajectory state and the real observation (detection):
\begin{equation}\label{eq:gating-cond}
J^{\rm RDA}_{ij} = \left(\boldsymbol{z}_t^j - \boldsymbol{H}\boldsymbol{x}_t^i  \right)^T \boldsymbol{S}^{-1}_t \left(\boldsymbol{z}_t^j - \boldsymbol{H}\boldsymbol{x}_t^i  \right), % = \boldsymbol{y}_t^{ijT} \boldsymbol{S}^{-1}_t \boldsymbol{y}_t^{ij},
\end{equation}
where \mbox{$\boldsymbol{z}_t^j - \boldsymbol{H}\boldsymbol{x}_t^i$} is the innovation process and \mbox{$\boldsymbol{S}_t$} its covariance matrix computed as \mbox{$\boldsymbol{H}\boldsymbol{P}_t^i\boldsymbol{H}^T + \boldsymbol{R}$}, and are both obtained as part of the KF update step.

The choice of two different score functions for RD and RDA processing is motivated by the different properties of the radar maps in the two cases. In the RDA space, trajectory tracking uses range and angle information, which leads to compact clusters around the centroids. Conversely, the velocity information that is used in the RD space leads to sparse clusters along this dimension, and the IOU score allows one to control the box shape, i.e., its form factor through $h_{B}$ and $w_{B}$, in order to weigh less a superposition along the velocity axis than that along the range axis. This significantly mitigates the tracking errors due to cluster sparsity in the RD space.

Given the cost matrix, the Hungarian algorithm \cite{kuhn1955hungarian} is used to efficiently obtain the associations yielding the lowest total cost, with complexity \mbox{$\mathcal{O}((K_{t-1} D_t)^3)$}. The algorithm uses the cost matrix as input and pairs each trajectory with only one cluster detection. 

\subsection{Trajectory management}
\label{sec:trackman}

During trajectory tracking we must deal with {\it (i)} undetected trajectories and new cluster detections (that is the case of a \mbox{non-square} matrix $\boldsymbol{J}$), {\it (ii)} trajectory instability due to missed detections, and {\it (iii)} presence of ghost targets generated by reflections from metal objects. To deal with these problems, we conceived the following trajectory management measures.

\subsubsection{Unmatched trajectories (RD and RDA)} 
All past trajectories that are not associated with any current cluster detection are marked as \textit{undetected} and are maintained for \mbox{$\texttt{max_age} = 10$} frames before being deleted. During these frames, their state is updated using \eq{eq:stateup}.  Cluster detections that are not associated with any existing trajectory are called \textit{unmatched}, and are initialized as new trajectories if they are detected for \mbox{$\texttt{min_det} = 15$} consecutive frames.

This mechanism makes the system robust to subjects that randomly appear on and disappear from the environment, and whose tracks are created or deleted as needed with an affordable delay. We stress that a subject reflection could go undetected, subsequently deleted and reinitialized for any reason, e.g., due to blockage at the beginning of the measurement sequence, and more generally, blockage at any point of the monitored sequence, or due to targets moving in and out of the scene. These temporary effects will not affect the correct tracking of the proposed system, whose trajectories will be continuously refined and reinitialized as soon as a reliable measurement is obtained.

\subsubsection{Ameliorating trajectory instability (RDA)} 
Trajectory instability and merging trajectories due to missed detections are a problem in the RDA case, where clutter is more significant. For this reason, we introduced a \emph{gating region} around each trajectory, i.e., a detection is never associated with the trajectory if the cost (squared Mahalanobis distance) of the association at time step $t$ is higher than a threshold value denoted by $\gamma$. This operation discards all the possible associations between a trajectory and clusters that lie outside of an ellipsoidal region whose shape and size are determined by the innovation covariance $\boldsymbol{S}_t$ and the threshold $\gamma$, which is typically chosen according to a desired level of confidence from an inverse $\chi^2$ distribution with $2$ degrees of freedom~\cite{bar2009probabilistic}. In this work, we use a $90$\% confidence, which leads to \mbox{$\gamma = 4.605$}.

\subsubsection{Dealing with merging trajectories (RDA)} 
Merging trajectories are detected by checking the Euclidean distance between their estimated states. If the distance between two trajectories gets lower than a minimum distance \mbox{$d_{\rm min} = 0.5$~m}, the trajectory with the highest variance in the last $5$ state estimations is deleted in order to favor stability.

\subsubsection{Removal of ``ghost'' targets (RDA)} 
As a last trajectory management measure, we eliminate all trajectories whose estimated state lies outside of the room boundaries.
This has a significant positive effect in removing \textit{ghost targets} due to multipath reflections on metal objects and wide flat surfaces. These unwanted reflections often closely resemble the direct ones from the real subjects, but appear at different angular positions, and at a longer distance due to the longer path followed by the signal.
The method used in this work for the removal of ghost targets uses some prior information about the dimensions of the room. In practice, a coarse knowledge is sufficient and the room dimensions along the $x$ and $y$ reference axes are enough. This data could be obtained from a planimetry, as we did, or via some \mbox{pre-processing} performed on the radar signal. The latter approach is left as a future research work.

\subsection{Computation of $\mu$D time series} \label{sec:mud-trunc}

The $\mu$D signature of each target is computed by selecting those points belonging to the cluster that is currently associated with her/his trajectory. This allows obtaining a separate signature for each subject. Such signature is inputted into a DCNN based classifier to perform identification, see \secref{sec:cnn_class}. For the computation of the $\mu$D vector in a given time step, the received power over the range (RD) or range and angle (RDA) dimensions is accumulated, producing vectors with dimension equal to the number of considered Doppler bins, $n_{\rm chn}$. Hence, these vectors are stacked over time and passed to the DCNN classifier as a spectrogram image. This image is the input $\boldsymbol{X}$ for the following identification block, see  \secref{sec:cnn_class}.

\subsection{Identification -- DCNN} 
\label{sec:cnn_class}

\begin{figure*}[t!]
	\begin{center}   
		\includegraphics[width=16cm]{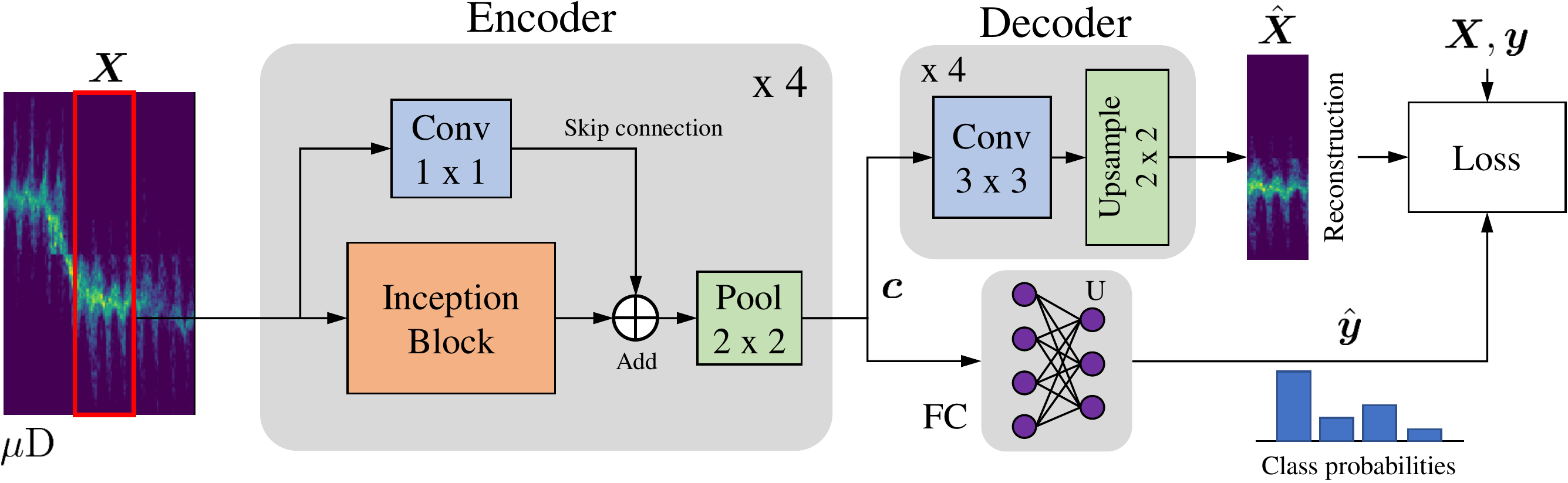} 
		\caption{Architecture of the proposed classifier.}
		\label{fig:prop-arch}
	\end{center}
\end{figure*}
\begin{figure}[t!]
	\begin{center}   
		\includegraphics[width=5cm]{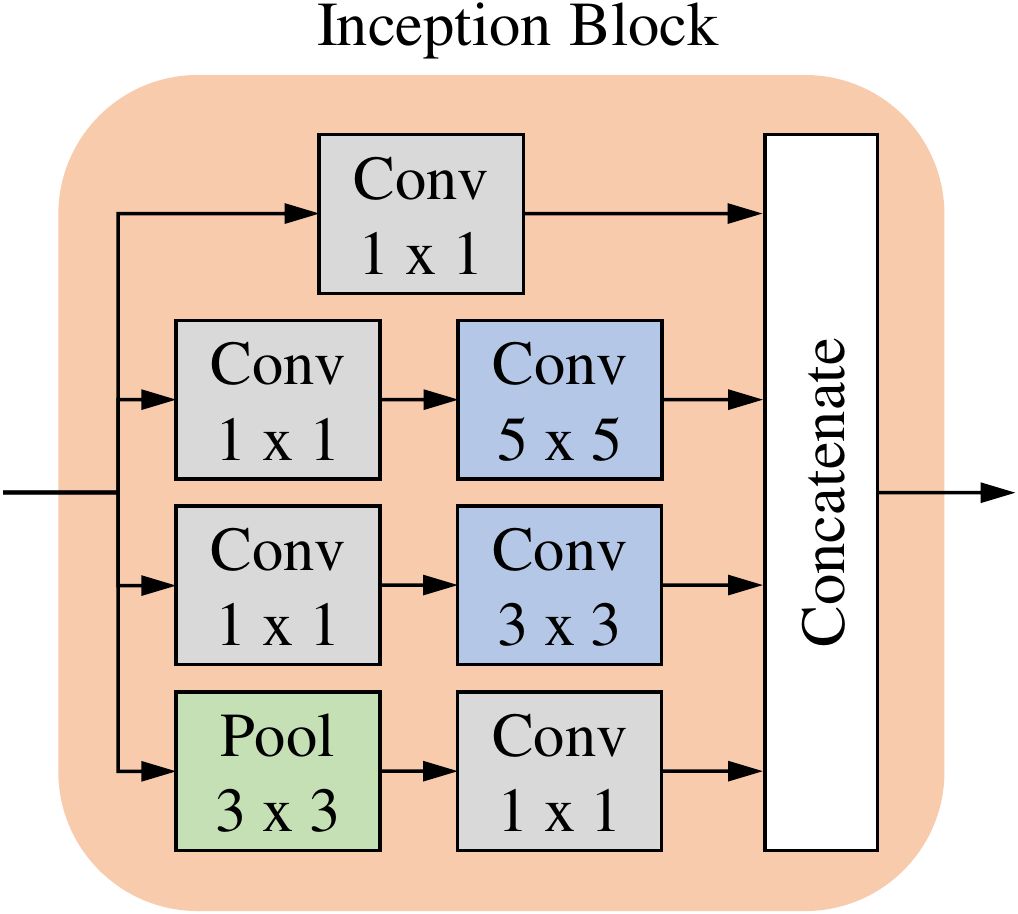} 
		\caption{Structure of the Inception Block.}
		\label{fig:blocks}
	\end{center}
\end{figure}

The proposed classifier architecture is a DCNN. This kind of neural network is suited for classification and feature extraction when the input data  exhibits spatial structure, like in image processing applications. The main components of the DCNN are convolutional layers, where the input is convolved with a filter (or \emph{kernel}) of learned weights in order to recognize certain patterns, organized into so called \emph{feature maps}, that become more and more complex and abstract with the depth of the layer. DCNNs have been broadly utilized in the last few years for feature extraction in spectrogram data, e.g., in speech recognition and audio processing applications~\cite{oord2016wavenet}.

The proposed DCNN is based on inception and residual networks structures, two architectures that are commonly used in \mbox{state-of-the-art} image classification tasks. IBs are a DCNN structure developed for complex feature extraction at different scales, using at every layer of the DCNN different kernel sizes, in a parallel fashion, and concatenating the resulting feature maps~\cite{szegedy2015going}. In our case, \mbox{$1\times 1$}, \mbox{$3 \times 3$} and \mbox{$5 \times 5$} kernel filters are used at each layer, to extract small and wide scale characteristics of the $\mu$D signature. An efficient implementation of the single inception block is shown in \fig{fig:blocks}: the top branch uses \mbox{$1 \times 1$} convolutions, extracting small scale features, the two following branches from the top use \mbox{$3\times 3$} and \mbox{$5 \times 5$} convolutions, which are preceded by \mbox{$1 \times 1$} convolutions to reduce their complexity, i.e., the number of feature maps, and prevent the number of parameters from becoming too large. The bottom branch performs a \mbox{$3\times 3$} max pooling operation, still extracting small scale features, but from a downsampled representation of the input.

Residual networks instead rely on skip connections between the input and the output of convolutional blocks~\cite{he2016deep}, in order to make the network learn the residual representation of the data with respect to the input. This has been shown to allow deeper networks to be trained faster and with significant performance gains. In our case, skip connections are placed between the input and the output of each IB, summing the respective tensors. A \mbox{$1 \times 1$} convolution is applied to each skip connection to adjust the number of feature maps, so that it matches that at the output.

The input signal $\boldsymbol{X}$ is a sequence of \mbox{$W_c=30$} frames of $\mu$D vectors, corresponding to \mbox{$W_c T_{\rm seq} = 2$}~seconds of measurement time for each subject. The number of Doppler bins that were selected is \mbox{$n_{\rm chn}=200$} (see \secref{sec:results} for a detailed description of the evaluation setup), so the input image has dimension \mbox{$200\times 30$}. The input $\boldsymbol{X}$ is passed through the three blocks composing the DCNN, namely, an {\it encoder}, a {\it decoder} and a {\it fully connected} (FC) network. The encoder network, $\mathcal{E}$, is composed of {\it four} stacked IBs with a number of output feature maps respectively equal to $16$, $32$, $64$ and $16$; the blocks are separated by \mbox{$2\times2$} max pooling layers, which perform dimensionality reduction.

The flattened output of the encoder,  $\boldsymbol{c}$, is a latent representation of the input with lower dimensionality, i.e., a {\it code}, and is fed to both the decoder and the FC network. In detail,  
\begin{enumerate}
\item the {\bf decoder network} $\mathcal{D}$ learns to reconstruct the input image. $\mathcal{D}$ is a CNN with four layers, \mbox{$3\times 3$} filters in each layer, and a number of feature maps respectively equal to $32$, $32$, $16$ and $1$. A \mbox{$2\times 2$} upsampling step is carried out before each convolution. The reconstructed copy of the input is called $\hat{\boldsymbol{X}}$. This branch of the classifier does not directly contribute to the classification result, but it is used during the training phase to guide the network towards extracting meaningful features, acting as a {\it regularizer}. To the best of our knowledge, the use of a decoder network for this class of problems is an original contribution of our design. We found its use to be effective, leading to accuracy improvements in the order of $2-3$\% in the test set.

\item The {\bf FC network} $\mathcal{F}$ outputs a $U$-dimensional vector containing the probabilities that the input belongs to each class using a \mbox{one-of-$U$} encoding, i.e., $\hat{\boldsymbol{y}} = [\hat{y}_1, \dots, \hat{y}_U]^T$, with $\hat{y}_u \in \left[0,1\right]$ and $\sum_{u} \hat{y}_u=1$. The network is composed of one hidden layer with $128$~neurons. ELU activation functions~\cite{clevert2015fast} connect the input to the hidden layer neurons, while a \texttt{SoftMax} layer is used to compute the $U$ output probabilities.
%with an hidden layer of  $128$~units.
%The output layer has $U$ neurons, it applies a softmax function and outputs the probabilities that the input belongs to each class in \emph{one-of-K} representation $\hat{\boldsymbol{y}} \in \left\{0,1\right\}^U$.
\end{enumerate}
The following equations formalize the input-output relations for the encoder, decoder and FC blocks
\begin{equation}
\boldsymbol{c} = \mathcal{E}(\boldsymbol{X}),\quad	\hat{\boldsymbol{X}} = \mathcal{D}(\boldsymbol{c}),\quad \hat{\boldsymbol{y}} = \mathcal{F}(\boldsymbol{c}).
\end{equation}
The loss function of the full architecture is a weighted sum of the loss function of the decoder, which measures the difference between the original input image $\boldsymbol{X}$ and the reconstructed one $\hat{\boldsymbol{X}}$, and the loss of the FC branch (classification). For the former, we choose the average \mbox{per-pixel} binary \mbox{cross-entropy} loss, while the categorical \mbox{cross-entropy} loss between the predicted and the true labels $\boldsymbol{y}$ is used for the latter. We call $n_{\boldsymbol{X}} = n_{\rm chn} W_c$ the number of elements in the $\mu$D input image, $U$ the number of classes (the known user identities) and $\alpha_{\rm rec}$ is a weighting factor. The $p$-th pixels of the input and reconstructed images, with values in $\left[0, 1\right]$, are denoted respectively by $X_p$ and $\hat{X}_p$ and the total weighted loss function is
\begin{multline}\label{eq:loss-function}
\mathcal{L}(\hat{\boldsymbol{X}}, \boldsymbol{X}, \boldsymbol{y}) = \underbrace{-\left(1-\alpha_{\rm rec}\right)\sum_{u=1}^{U}y_u \log(\hat{y}_u)}_{\text{Classification branch term}}- \\-   \underbrace{\frac{\alpha_{\rm rec}}{n_{\boldsymbol{X}}} \sum_{p=1}^{n_{\boldsymbol{X}}}X_p \log(\hat{X}_p) + (1-X_p) \log(1-\hat{X}_p)}_{\text{Reconstruction branch term}}.
\end{multline}
\fig{fig:prop-arch} shows the complete structure of the classifier. 
As a regularization measure, after each layer of the encoder and the FC branch we apply batch normalization~\cite{ioffe2015batch}.
All the hidden nodes in the network use the ELU activation function~\cite{clevert2015fast}.
The complete neural network has $560,819$ tunable parameters ({\it network size}).

\subsection{Labeling and trajectory correction procedure}
\label{sec:correction}

Previous approaches to human identification from \mbox{mm-waves} obtain trajectories rely on the sole KF output, for the entire movement and, in a following step, perform the classification on the \mbox{pre-computed} trajectories using, e.g., a neural network of some kind \cite{zhao2019mid}. Now, consider the trajectories of two users $1$ and $2$ that, at a certain point in time, intersect in the considered RD/RDA space. At this point, the two users cannot be distinguished, as their clusters largely overlap, and the trajectories are tracked again by the KF from the moment in which their clusters set apart. The target association procedure, however, beyond this point, may wrongly associate trajectories with detections, i.e., assigning trajectory $1$ to user $2$ and \mbox{vice-versa}. This problem can be hardly corrected with previous algorithms, whereas it is solved with the interactive procedure that we designed, and that we detail in this section. With our technique, identities are obtained in an online manner. Moreover, although  the association of trajectories to clusters (see \secref{sec:hungarian}) may be erroneous, due to the overlap of the user clusters, as soon as the trajectories set apart again, the association is corrected using the output of the DCNN classifier. Note that this is not possible by solely exploiting the KF, as its memory amounts to a single time step, which is insufficient to solve this issue. Next, the procedure is formally described.  

Applying the classifier to the $\mu$D signatures from the $K_t$ current trajectories, returns $K_t$ $U$-dimensional vectors, which contain the probabilities that each trajectory belongs to one of the $U$ (known) user classes. 
Hence, we build a \mbox{$K_t \times U$} matrix, $\boldsymbol{\Gamma}_t$, by stacking these vectors. The matrix contains in position $(i, j)$ the probability that trajectory $i$ belongs to subject $j \in \left\{1, \dots, U\right\}$. Following a reasoning similar to that in \secref{sec:hungarian}, we can interpret the matrix $\boldsymbol{\Gamma}_t$ as a {\it score} matrix for the associations between trajectories and classes. Therefore, the optimal assignment of the labels in the current time step is obtained by applying again the Hungarian algorithm on $-\boldsymbol{\Gamma}_t$, which represents the association costs. This approach makes it possible to {\it jointly} label all the trajectories. From the properties of the Hungarian algorithm, it descends that the same class is never assigned to more than one subject. A subject is classified as {\it unknown} in case no label is assigned to her/him by the algorithm (which happens if \mbox{$K_t > U$}) or when the score outputted by the DCNN is lower than $0.1$ (a threshold that we set to avoid low probability associations).
%Hence, we build a \mbox{$K_t \times U$} score matrix by stacking these vectors and apply again the Hungarian algorithm to jointly obtain the best assignment across all trajectories. From the properties of the Hungarian algorithm, it descends that the same class is never assigned to more than one subject. A subject is classified as {\it unknown} in case no label is assigned to her/him by the algorithm (which happens if \mbox{$K_t > U$}) or when the score outputted by the DCNN is lower than $0.1$ (a threshold that we set to avoid low probability associations).

To enhance the stability in the identification process, the current labels that are outputted at time $t$ by the DCNN are used with the past ones as follows. 
%Instead of using instantaneous labels obtained as above to classify the targets, we act as follows
\begin{itemize}
\item for each trajectory, we store the past labels that are outputted by the DCNN in a list;
\item at \mbox{$t = 0$}, subjects are identified using the instantaneous labels, as no past information is available;
\item at time step \mbox{$t > 0$}, each trajectory $i$ is classified considering the most recent $W_h$ labels that are outputted by the DCNN classifier up to and including time $t$, i.e., at time steps $(t-W_h+1), \dots, (t-1), t$. If all these $W_h$ labels match, we assign their common value to the trajectory; this will be the final identity label that is outputted at time $t$. If instead different values appear in this list, we keep the final label that was previously assigned, at time $(t-1)$, to trajectory $i$. Note that, in case the $W_h$ values in the list for any trajectory $i$ differ, the procedure will maintain the previous label until the DCNN will output a sequence of $W_h$ matching labels. 
\end{itemize}
We remark that the value of $W_h$ encodes the level of {\it temporal stability} that is required to accept a change in the identity that is outputted by the algorithm, for any trajectory. In fact, this procedure introduces additional stability in the identification, as misclassifications that only last a few time steps are avoided. A cost is however paid in terms of correction speed when a tracking error occurs, e.g., when trajectories are swapped between users. As such, a desirable tradeoff has to be identified between the stability in the identification results (large $W_h$) and the delay in compensating for tracking errors (small $W_h$).

\section{Experimental results}\label{sec:results}
\subsection{Measurement setup and parameters} \label{sec:setup}
\begin{table}[t!] 
	\begin{center}
		\begin{tabular}{lcr}
			\toprule
			\multicolumn{3}{c}{{\bf Measurement parameters }} \\
			\midrule
			Antenna el. spacing & $\delta$ & $1.948$~mm\\
			Number of receiving antennas & $L$ & $16$ \\ 
			Start frequency & $f_o$		& $76$~GHz      \\
			Chirp bandwidth & $B$		& $2$~GHz     \\
			Chirp duration & $T$       		& $180$~$\mu$s   \\
			Chirp repetition time  & $T_{\rm rep}$ & $250$ $\mu$s\\
			No. samples per chirp & $N$    & $512$ \\
			No. chirps per seq. & $P$  & $256$ \\
			Frame rate & $1/\Delta t$       & $15$~fps    \\
			ADC sampling frequency & $F_s$          & $2.857$~MHz       \\
			Range resolution & $\Delta R$    & $7.5$~cm       \\
			Velocity resolution & $\Delta v$ & $3.040$~cm/s      \\
			%			Max. range & $R_{\rm max}$     &   54 m  \\
			%			Max. velocity & $v_{\rm max}$ &   3.896  m/s   \\
			\bottomrule
		\end{tabular} 		
	\end{center}
	\caption{Summary of the radar working parameters used in the evaluation session. \label{tab:radar-params}}
\end{table}
\begin{figure*}[t!]
	\begin{center}   
		\centering
		\subcaptionbox{Map\label{fig:map}}{\includegraphics[width=11cm]{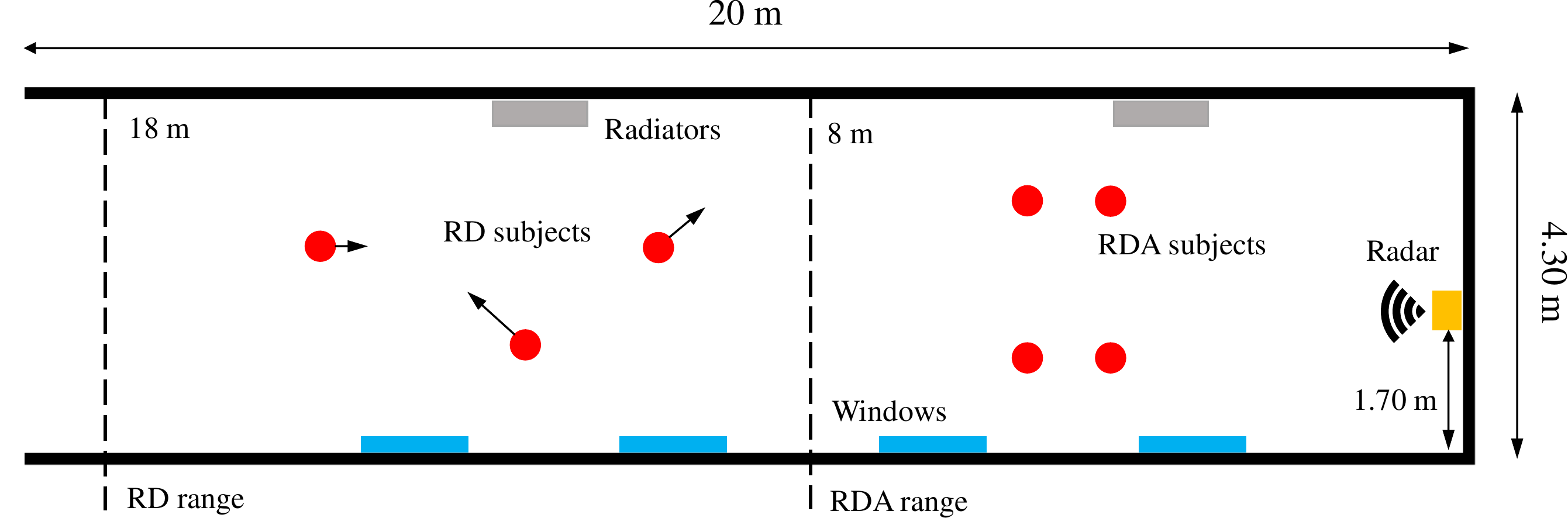} }
		\subcaptionbox{Camera view\label{fig:meas-image}}{\includegraphics[width=4.8cm]{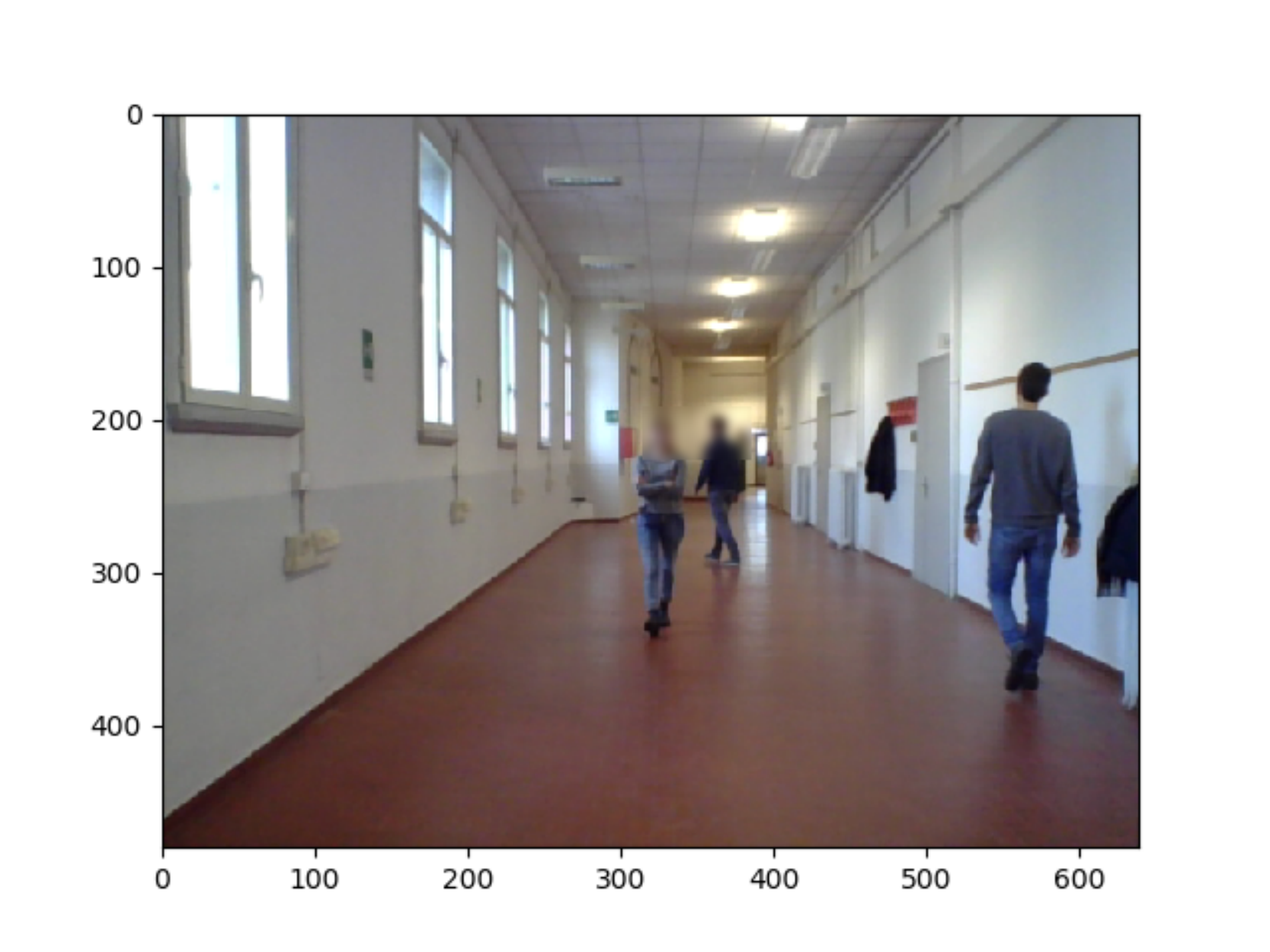}}
		\caption{Measurement room A.}
		\label{fig:meas-room}
	\end{center}
\end{figure*}

The proposed framework is evaluated using an INRAS RadarLog device working at $77$~GHz center frequency. The \mbox{front-end} features $2$ transmitting antennas and $16$ receiving antennas organized as a linear array. The device working parameters are set up as in \tab{tab:radar-params}. Operating in LFMCW mode, we respectively utilize one transmitting and $L=16$ receiving antennas.
%Operating in LFMCW mode we can exploit the $2$ transmitter antennas in time division multiplexing (TDM) to fully utilize the MIMO capabilities forming a virtual receiver array of $32$ elements. However, in the following we limit ourselves to use one transmitting antenna, exploiting $16$ receiving channels. In this way, the frame sampling rate is not halved by the TDM scheme, which would reduce the time resolution of the $\mu$D signatures, at the cost of an affordable reduced resolution in the AoA processing. 
To obtain ground truth values for the \mbox{multi-target} measurements we used a camera which was \mbox{time-synchronized} with the radar device: the resulting video was used to identify and track the users within the indoor space.

To thoroughly evaluate the proposed system, several measurement campaigns were conducted, using two measurement rooms with very different dimensions, shapes and propagation environments.

Measurement room A is a $4.3\times20$~meters corridor, where the radar was positioned on the short edge as depicted in \fig{fig:meas-room}. The presence of several large windows and some radiators that become sources of unwanted reflections and ghost targets makes our evaluation room very challenging. Room A was the main environment considered in this work, where the majority of tests was carried out, along with the collection of the training data for the classifier.

Room B is a $8\times4$~meters research laboratory, with furniture and devices left in place in order to mimic a \mbox{real-life} indoor scenario. In addition, other people not involved in the measurements were also present in the room, outside the tracking area, further increasing the difficulty of tracking and identification tasks. Room B was utilized to validate our approach and investigate the generalization capabilities of the classifier, which was only trained on data from room A.

We collected radar data for the training and validation of our algorithm for the following experiments.
\begin{enumerate}
	\item {\bf Training the classifier on single subjects (room A).} We collected RDA data from 4 different subjects (S1, S2, S3 and S4) with ages ranging from $24$ to $31$ years and different body shapes and weights.
	Each subject was asked to walk alone, freely and without any restrictions within the measurement room for around $22$ minutes, to collect $20$ sequences of $500$ frames, for a total of $10$~thousand frames per subject. The sequences were acquired in two different days to reduce the effects of clothing or physical conditions. 
	
	\item {\bf Evaluating the performance of RD multi-person identification (room A).} We acquired $4$ test sequences of $1,250$ \mbox{RD-only} frames, $2$ of them with $2$ targets (S1 and S2) and the other $2$ with $3$ targets (S1, S2 and S3). All subjects were asked to walk freely, without space constraints and varying their walking speed.
	
	\item {\bf Evaluating the performance of RDA \mbox{multi-person} identification (room A).}  We acquired $6$ test sequences of $500$ RDA frames with $4$ targets. To make the test more challenging, we had the targets walking in a \mbox{square-like} fashion, with the first two subjects and the second two being at the same distance from the radar device, and with a small distance of about $1$~meter between the two pairs, as shown in \fig{fig:meas-room}. All targets are constrained to walk at (approximately) the same speed. This setup has been intentionally selected, as it represents a \mbox{worst-case} for the \mbox{RDA-based} system. In fact, in this case subjects can not be distinguished by their different velocities or their range, and the detection/tracking has to mainly rely on the angular information, which is less accurate than the range or the velocity\footnote{The angular resolution degrades as the angle of arrival of the reflections approaches $\pm\pi/2$, see \eq{eq:angle-est}.}. Moreover, the classifier is forced to make the identity decision based on the features of the $\mu$D spectrogram that encode the way of walking of the subjects, as their speed is the same. We stress that this type of analysis is new: often, $\mu$D classifiers based on neural networks include the \mbox{non-informative} average walking speed as a discriminative feature, leading to poor accuracy when subjects have similar velocities, e.g.,~\cite{vandersmissen2018indoor}.
	\item {\bf Validating the performance in a different measurement room (room B).} We collected RDA data with two subjects in a different environment in order to asses whether the system, in particular the DCNN classifier, generalizes to an unseen domain. No data from room B is used during the training of the classifier. The $2$ sequences obtained in room B contain $500$ frames each, and the walking patterns of the subjects were constrained similarly to point $3$). The room boundary values, see \secref{sec:trackman}, were modified complying with the new room dimensions to effectively deal with ghost targets. Note that only two subjects were involved in the experiments for Room B as the walking space available was reduced with respect to Room A. However, the density of users per square meter was higher for Room B, which again corresponds to a more challenging setup.
\end{enumerate}

With the considered parameters, raw radar frames have a shape of \mbox{$N \times L \times P = 512\times 16 \times 256$} points along the fast-time, antenna element, and slow time dimensions respectively. We used $64$ points for the DFT along the angle dimension and $256$ points for DFT along Doppler dimension. For the range dimension, we used $1,024$ points for the DFT, extracting ranges from $0$ to $10$~m for RDA data ($253$ bins) and from $0$ to $18$~m in case of RD maps ($497$ bins).
The contributions due to static objects were removed by cutting the $8$ central Doppler channels, corresponding to velocities in the range $[-0.138,0.138]$~m/s, and the first and last $24$ channels corresponding to velocities outside the interval $[-3.160, 3.160]$~m/s were also removed as they did not contain any useful information. The resulting radar maps after DFT processing have dimension \mbox{$253 \times 64 \times 200$} points for RDA and $497 \times 200$ points for RD, which corresponds to a $34$-fold increase in the data frame size for RDA with respect to RD. We performed $\mu$D extraction by summing over the range and angular dimensions, obtaining spectrograms with $200$ Doppler bins and variable time length depending on the sequence ($500$ or $1,250$ frames).

When using \mbox{range-Doppler} DFT processing, a common assumption is that the target covers a smaller distance than the length of a range bin during a single frame period, i.e., $vPT_{\rm rep} < \Delta R$, with $v$ being the velocity of the subject. If this assumption is not satisfied, the echo from the target spreads across multiple range bins (\textit{range migration})~\cite{barrick1973fmcw,richards2010principles}.
However, this effect is not harmful in our case, where the typical extension of human subjects along the range dimension goes from $0.5$~m up to $1.5$~m, which correspond to $6$-$20$ range bins. Indeed, the average walking speed of human subjects in indoor environments is around $1.5$~m/s~\cite{willen2013walking}, which lead to a ranging accuracy reduction of at most $1-2$ range bins, which is deemed negligible with respect to the target size.

\subsection{Training phase}
\label{sec:training}

We implemented the classifier network using TensorFlow $2.0$ and the Keras API. Training was performed on a NVIDIA RTX $2080$ GPU with $8$~GB of RAM.

The $20$ $\mu$D sequences per target obtained from the measurements in room A were split into windows of $30$ frames along the time dimension, with an overlap of $25$ frames. The resulting images were divided into training and validation sets, $90$\% and $10$\% of the images respectively, and testing was carried out on the \mbox{multi-target} sequences. Data augmentation was applied to enlarge the training set: for each training image we generated $4$ additional images by 
\begin{enumerate}
	\item adding Gaussian noise with zero mean and variance $0.05$,
	\item setting to zero pixels in the image with a probability of $0.3$ ({\it random corruption}),
	\item setting to zero $8$ adjacent columns (time frames) starting from an index selected uniformly at random ({\it time masking}),
	\item setting to zero $20$ adjacent Doppler bins starting from an index selected uniformly at random ({\it frequency masking}).
\end{enumerate}
These images were used as input $\boldsymbol{X}$ of the encoder, setting the reconstruction target $\hat{\boldsymbol{X}}$ at the output of the decoder to be the original image, to force the \mbox{encoder-decoder} pair to learn key structural properties of the input (the same strategy is exploited to train denoising \mbox{auto-encoders} (DAE)~\cite{vincent2010stacked}).
The model was trained on the training set until convergence of the loss $\mathcal{L}(\hat{\boldsymbol{X}}, \boldsymbol{X}, \boldsymbol{y})$ in \eq{eq:loss-function} on the validation set, using the Stochastic Gradient Descent (SGD) optimizer with Nesterov momentum $0.95$ and \mbox{$\alpha_{\rm rec}=0.6$}.
The learning rate was adaptively lowered by a factor of $0.5$ when the validation loss was not improving for more than $5$ consecutive epochs, from an initial value \mbox{$\eta = 5 \cdot10^{-3}$}. We applied $L_2$ regularization with coefficient \mbox{$\lambda = 3 \cdot 10^{-3}$} on the network weights and dropout with probability \mbox{$p_{\rm drop}=0.5$} for the fully connected layers, to reduce overfitting on the training data.

In terms of inference time, the proposed DCNN takes on average $24$~ms to perform the classification of a $2$~seconds long $\mu$D input. The $\mu$D signatures of all the tracked targets are fed to the network in a single batch so that only one \textit{forward pass} is performed in a time step.

\subsection{DCNN evaluation on the IDRad dataset (single-target)}\label{sec:idrad}

\begin{table}[t!] 
	\begin{center}
		\begin{tabular}{lc}
			\toprule
			\textbf{Classifier} & \textbf{Accuracy (IDRad) \%}\\
			\midrule
			DCNN \cite{vandersmissen2018indoor} & 78.46\\
			RCN \cite{jalalvand2019radar} & 75.65\\
			SIN + LSTM \cite{polfliet2018structured}& 89.56\\
			DCNN with IBs (our approach)& 90.69\\			
			\bottomrule
		\end{tabular} 		
	\end{center}
	\caption{Comparison between the proposed classifier and available benchmarks from the literature on the IDRad test set. 	\label{tab:idrad-conf}}
\end{table}

As a first evaluation phase, we trained and validated the proposed DCNN on {\it IDRad}\footnote{https://www.imec-int.com/en/IDRad}, a publicly available dataset of $77$~GHz radar $\mu$D signatures~\cite{vandersmissen2018indoor}. The dataset contains RD frames from $5$ different subjects walking one at a time in the environment and hence, \mbox{multi-target} identification is not possible using this dataset. Training and validation/test data are collected in two different rooms.

Using the IDRad dataset, we have assessed the performance of our framework for the single person identification problem and have compared it with available benchmarks~\cite{vandersmissen2018indoor,polfliet2018structured,jalalvand2019radar}. For a fair comparison against previous work, we adapted the DCNN to accept as input $\mu$D sequences with length of $45$ frames instead of $30$. We found that our classifier generalizes well, with an overall average accuracy of $90.69$\%, with slight variations across different targets, but always above $88$\%. The comparison between the performance of our approach and the schemes in the literature is presented in~\tab{tab:idrad-conf}. Our classifier is the most accurate, significantly outperforming the previous DCNN approach~\cite{vandersmissen2018indoor}, the one based on reservoir computing networks (RCN) \cite{jalalvand2019radar}, and performs slightly better than~\cite{polfliet2018structured}, where a structured inference network (SIN) and \mbox{long-short} term memory recurrent neural networks (LSTM) are used. We believe this improvement is achieved due to the use of IBs, which allow for feature extraction at different scales, without significantly increasing the network complexity, which would easily lead to overfitting.

\subsection{Performance metrics} \label{sec:metrics}

\begin{table*}[t!] 
	\begin{center}
		\begin{tabular}{lcccccccccccc}
			\toprule
			&\multicolumn{7}{c}{{\bf Range-Doppler}}& \multicolumn{5}{c}{{\bf Range-Doppler-Azimuth}} \\
			\cmidrule(lr){2-8}\cmidrule(lr){9-13}
			&\multicolumn{3}{c}{ 2 Subjects}&\multicolumn{4}{c}{3 Subjects}& \multicolumn{5}{c}{4 Subjects}\\
			&S1&S2                 &      {\bf Avg.}          &S1&S2&S3    &  {\bf Avg.}& S1&S2&S3&S4&{\bf Avg.} \\
			\cmidrule(lr){2-4}\cmidrule(lr){5-8}\cmidrule(lr){9-13}
			Accuracy \%  &98.24&97.69&97.96&95.75&98.65&91.38&95.26&99.52&98.26&100.0&95.56&98.27\\
			$r_{\rm und}$ \%&0&0&0&6.65&27.31&0&11.32&6.51&6.17&18.64&6.08&9.35\\
			$r_{\rm unk}$ \%&4.54&2.53&3.54&0.75&2.79&9.51&4.34&0&0&0&0&0\\
			\bottomrule
		\end{tabular} 		
	\end{center}
	\caption{RD and RDA average performance over the test sequences from room A ($W_h = 9$). 	\label{tab:results}}
\end{table*}

To train and test the proposed processing pipeline in a \mbox{multi-target} setting, we have collected our own RD and RDA data across several measurement campaigns (see \secref{sec:setup}). The performance of the final classifier are evaluated in terms of 
{\it (i)}~ \textbf{accuracy}, i.e., the ratio between the number of frames in which the target is correctly identified and the number of frames in which it is detected and assigned a label different from {\it unknown} (see \secref{sec:correction}); 
{\it (ii)}~the \textbf{undetected ratio} ($r_{\rm und}$), i.e., the ratio between the number of frames in which a target is undetected\footnote{A target is said to be {\it undetected} if the number of consecutive missed detections is sufficient to eliminate its trajectory from those that are being tracked by the algorithm. As such, the target is no longer identified.} and the total number of frames collected;
{\it (iii)}~the \textbf{unknown ratio} ($r_{\rm unk}$), the ratio between the number of frames in which the target is labeled as {\it unknown} and the total number of frames collected. This last metric is a measure of the uncertainty of the identification framework in providing a classification for the targets.

\subsection{Results for the RD signal (multi-target)}\label{sec:perf-rd}

In \tab{tab:results}, we report the results per subject using the metrics of \secref{sec:metrics}, averaged over the test sequences. In the evaluation, we discard the initial phase where the trajectories need to accumulate $30$ frames of $\mu$D data in order to provide the first image to the DCNN classifier.

With RD maps, the two targets case achieves the highest accuracy, with an average of $97.96$\%. With three targets, $r_{\rm und}$ increases for some subjects, as one may expect: having more targets in the same area leads to a higher probability of superposition of their clusters. In this case, the reflection coming from target $2$ is undetectable due to the fact that $27$\% of the frames for this user overlap with those of other users in the RD space (as they have a similar range and speed). An interesting point, however, is that the identification accuracy and $r_{\rm unk}$ are not significantly impacted with respect to the two targets case, meaning that the identification framework can recover from missed detections, still providing high accuracy when targets become detectable again.

A detailed analysis of the errors revealed that the main problem with RD processing is the {\it superposition of clusters} in the RD space: this occurs when subjects have similar range and speed, likely being detected as a single cluster. This is an intrinsic limitation of the RD space, and is not influenced by any of the system parameters. However, thanks to the proposed processing method, that allows \mbox{re-establishing} trajectories once clusters separate, and to correct errors using the identification outcomes (see \secref{sec:correction}), the system still provides correct results for a very high percentage of time. Other techniques from the literature treat tracking and identification separately, and are therefore unable to deal with \mbox{multi-target} RD identification because of their inability of recovering from erroneous tracking.

\begin{figure}[t!]
	\begin{center}   
		\includegraphics[width=8.5cm]{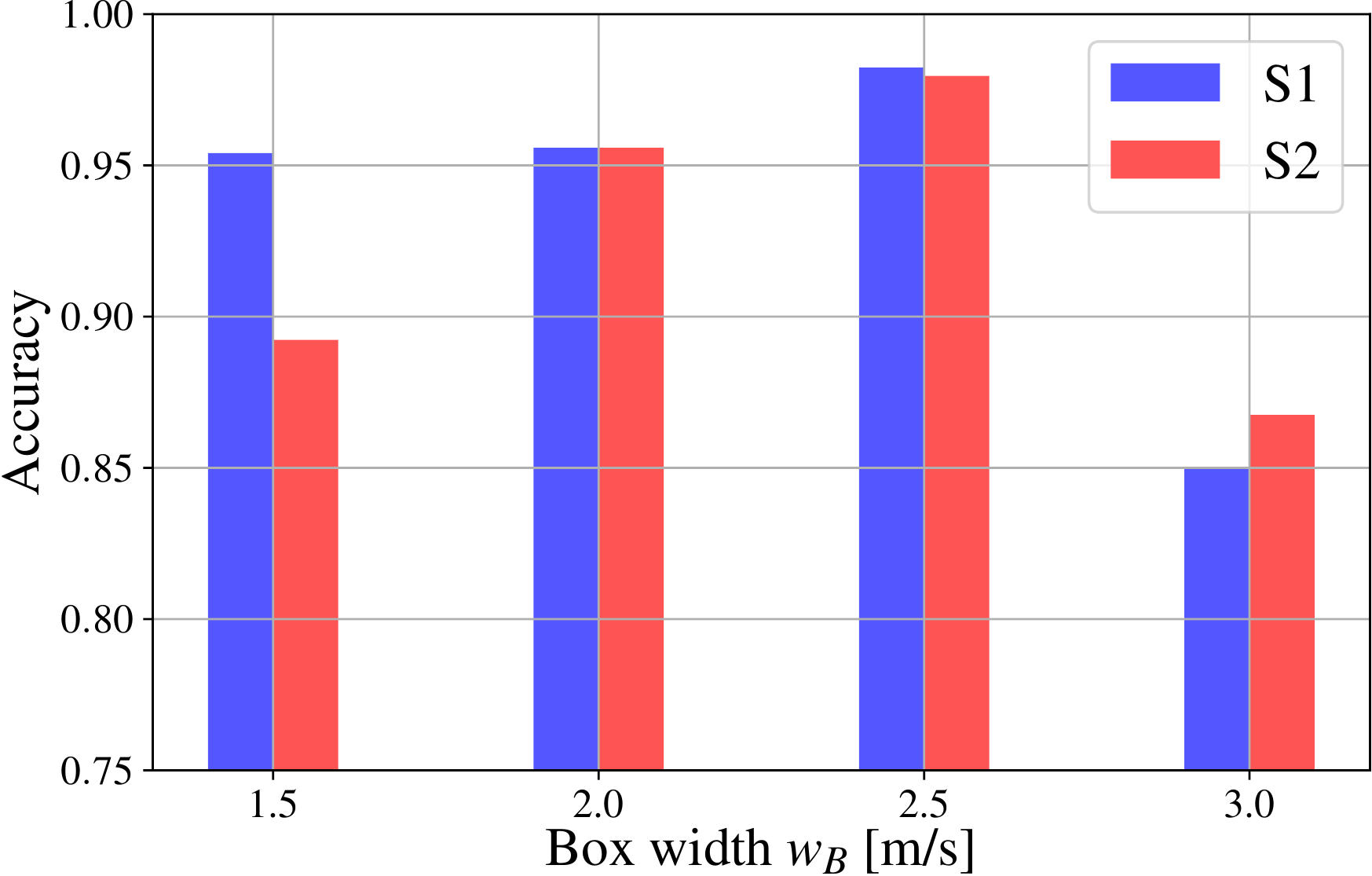} 
		\caption{Accuracy of RD identification by varying the box width $w_B$ along the Doppler dimension for two subjects, S1 and S2.}
		\label{fig:box-width}
	\end{center}
\end{figure}

As a last result, in \fig{fig:box-width} we show the impact of changing the box dimension along the Doppler axis, $w_B$, averaging the accuracy obtained on two targets. As expected, there is a \mbox{trade-off} between capturing most of the target's Doppler information (large $w_B$) and avoiding unnecessary overlap between boxes (small $w_B$), which may lead to classification errors. The chosen value for the results of \tab{tab:results} is $2.5$~m/s, as it provided the highest accuracy. The dimension of the box along the range dimension, $h_B$, is instead kept fixed at $2$~m.

\subsection{Results for the RDA signal (multi-target)}\label{sec:perf-rda}

\tab{tab:results} shows the results of RDA processing averaged over the $6$ test sequences: our system achieves an accuracy of $98.27$\% over $4$ targets. We recall that the initial phase in which the DCNN has to collect the first $30$ $\mu$D vectors is neglected in the computation, and only frames after this initial transient period are considered, as for the RD analysis.

%Given the dimensions of the considered environment, that corresponds to a density of $0.1$ person/m$^2$.
The relatively high people density ($0.1$ person/m$^2$) with respect to that in the RD analysis causes blockage to become more frequent, i.e., some subjects block the signal path to other targets during some frames, which explains the \mbox{non-negligible} average $r_{\rm und}$ of $9.35$\%. Conversely, $r_{\rm unk}$ is always zero for all subjects and all sequences, meaning that once a target is detected, the network has always enough data and confidence to produce a classification result. Remarkably, although $r_{\rm und}$ is greater than zero for all subjects, the identification accuracy is still very high (see in particular S3), which confirms once again the framework's ability to recover from missed detections. This is possible thanks to the correction algorithm of \secref{sec:correction}.

\subsection{Impact of training parameters}\label{sec:train-eval}
The proposed DCNN architecture, in terms of number of inception blocks, number of FC layers and presence of skip connections, was obtained using a greedy search procedure on the hyperparameters, i.e., we repeated the model training by changing one parameter at a time and selecting the value that led to the minimum loss. This procedure is suboptimal with respect to an exhaustive search or to Bayesian methods, but was preferred because of its lower computational cost. In the following, we focus on the most interesting and influential parameters, namely the value of $\alpha_{\rm rec}$ and the use of data augmentation, analyzing their impact on the final test accuracy.
	\begin{figure}[t!]
	\begin{center}   
		\includegraphics[width=8.5cm]{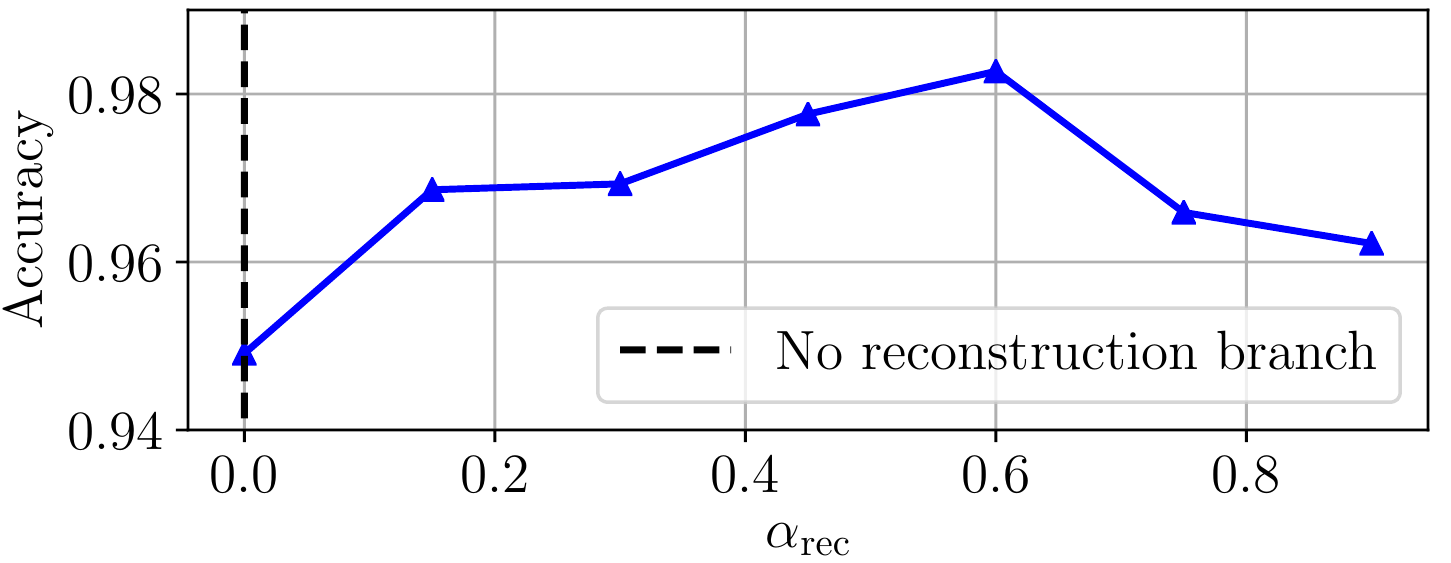} 
		\caption{Accuracy on the RDA test data depending on the reconstruction weight $\alpha_{\rm rec}$ (room A).}
		\label{fig:alpha-res}
	\end{center}
\end{figure}
The reconstruction weight \mbox{$\alpha_{\rm rec} \in [0, 1]$} tunes the relative importance of the classification and the reconstruction branches in the training loss. Using the reconstruction branch has shown to yield a slight improvement in the generalization capabilities of the DCNN, similarly to what is commonly achieved using regularization. Specifically, in \fig{fig:alpha-res} we plot the accuracy values obtained varying $\alpha_{\rm rec}$ from $0$, i.e., the reconstruction branch is disabled, to $0.9$. The best result is obtained for \mbox{$\alpha_{\rm rec} = 0.6$}. Remarkably, these improvements are only possible when using the \mbox{encoder-decoder} structure in combination with the data augmentation strategy described in \secref{sec:training}. Indeed, with no data augmentation (e.g., noise addition and random signal deletion/corruption) the reconstruction at the output of the \mbox{auto-encoder} becomes much easier and the feature extraction is less effective in capturing the true signal manifold. As a result, no major benefit is observed. 

In \tab{tab:train-frac}, we show the effect of only using a portion of the total training data available on the time required to  complete the training, and on the accuracy on the RDA test data. While the training time increases almost linearly when using $25$\%, $50$\%, $75$\% or $100$\% of the full training set ($22$~minutes per subject), the accuracy shows the smallest improvement when going from $75$\% to $100$\%. This is a saturation effect on the model's performance that is customary with neural network training. In the last column of \tab{tab:train-frac}, we show the results of training the model on the whole available dataset without exploiting data augmentation. The accuracy decreases significantly, motivating the use of the augmentation techniques described in \secref{sec:training}.

\begin{table}[t!] 
	\begin{center}
		\begin{tabular}{lccccc}
			\toprule

			  \textbf{Training set \% used}& $25$\% & $50$\% & $75$\% & $100$\%  & No augm.\\
			  \midrule
			 Acc. \% (RDA 4 trg.) & 84.08 & 88.97 & 95.45 & 98.27& 89.97\\
			Training time [min.]& 8 & 15 & 23 & 30 & 9\\
			\bottomrule
		\end{tabular} 		
	\end{center}
	\caption{Impact of training set size and data augmentation on accuracy (RDA with four targets) and training time.}	
	\label{tab:train-frac}
\end{table}

\subsection{Integrated {\it vs} separate tracking and identification}

\begin{figure}[t!]
	\begin{center}   
		\includegraphics[width=8.4cm]{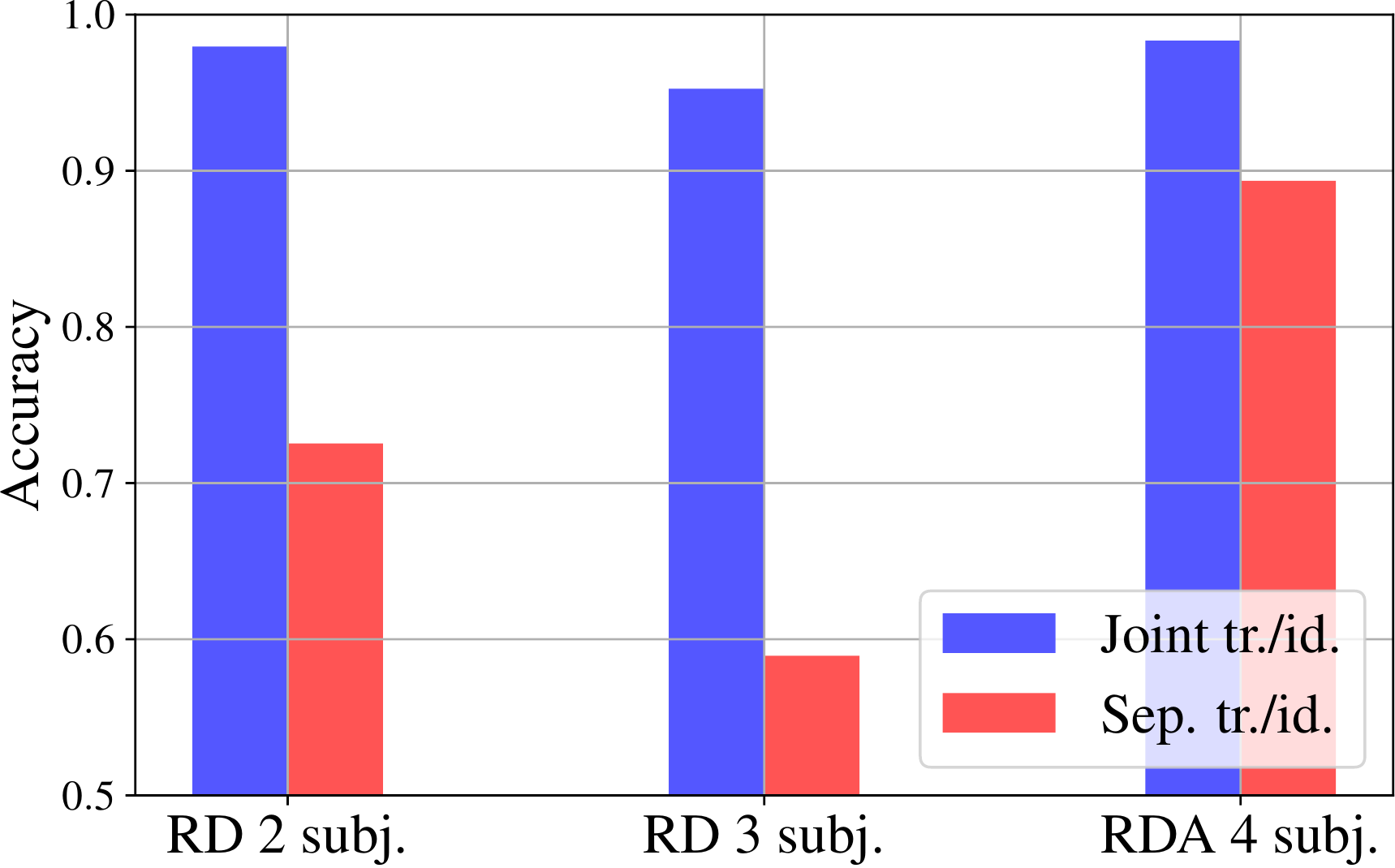} 
		\caption{Accuracy comparison between \textit{joint} (our approach) and \textit{separate} (previous work) tracking ({\it tr.}) and identification (\it{id.}).}
		\label{fig:feedback-comp}
	\end{center}
\end{figure}

As described in \secref{sec:correction}, the proposed system jointly performs tracking and identification. To quantify the improvement of this design with respect to separately obtaining trajectories and identities, we quantify the difference in the average accuracy when applying the two approaches (joint {\it vs} separate processing) on all the considered subjects and RD/RDA test sequences. \fig{fig:feedback-comp} confirms that our integrated approach is of key importance to enable precise RD identification, with improvements of $36.32$\% and $25.42$\% on the $3$ and $2$ subjects cases, respectively. For RDA processing, the improvement is smaller ($8.91$\%), due to the higher detection capabilities of the system in the RDA space, which makes cluster superposition and subsequent tracking errors less frequent. The improvement is however \mbox{non-negligible} and the proposed combined architecture is still very effective.

\subsection{Dimensioning the classification window $W_h$}

\begin{figure}[t!]
	\begin{center}   
		\includegraphics[width=8.7cm]{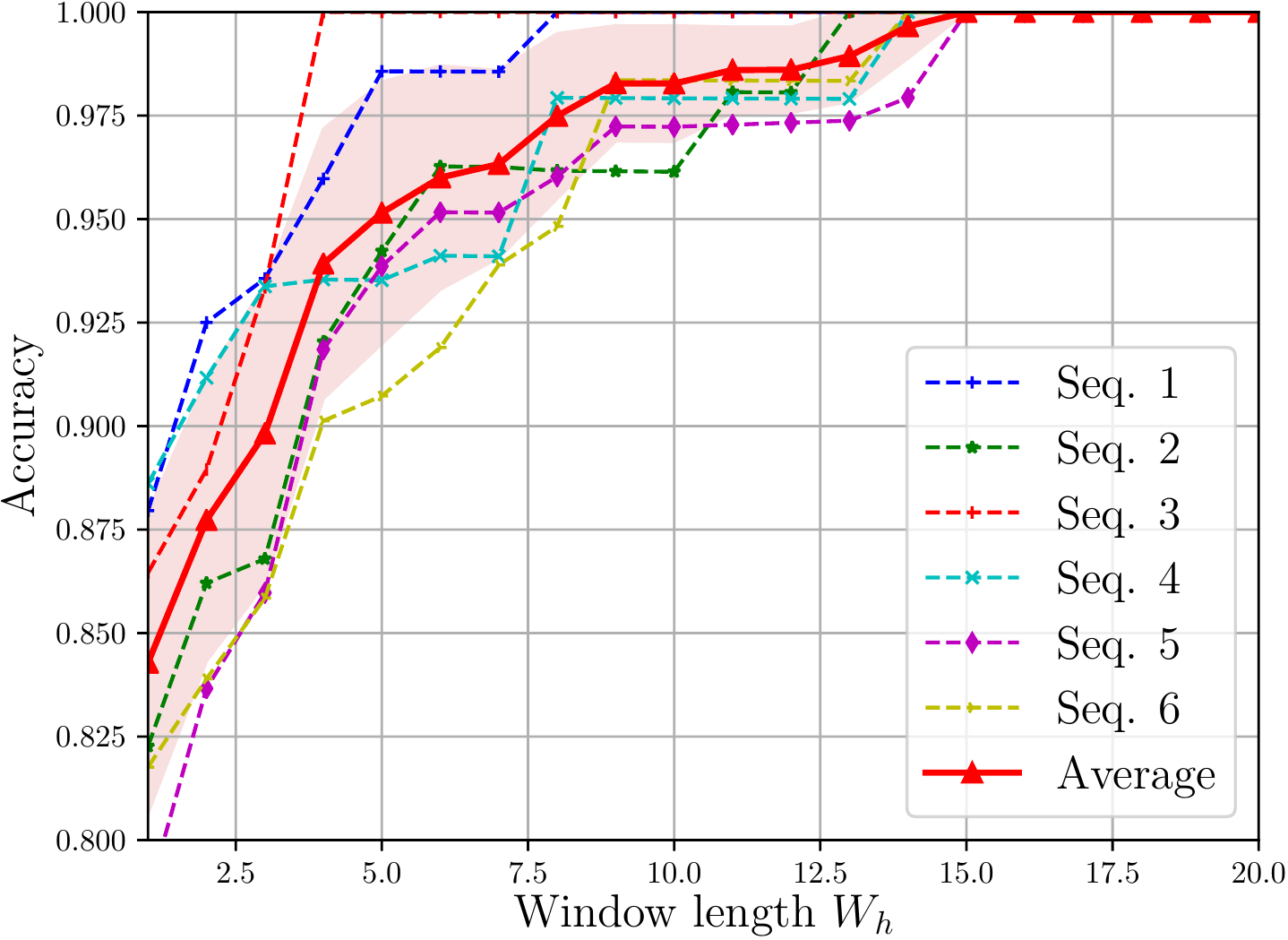} 
		\caption{Accuracy of the identification in RDA processing with respect to the length (i.e., number of frames) of the classification window $W_h$ (in room A). The red solid curve represents the average over all sequences, with uncertainty in terms of one standard deviation (shaded area).}
		\label{fig:rda-winacc}
	\end{center}
\end{figure}

As anticipated in \secref{sec:correction}, the classification window parameter $W_h$ plays an important role in the \mbox{trade-off} between online classification accuracy and speed in recovering from errors. In \fig{fig:rda-winacc}, we show the effect of varying $W_h$ from $1$ to $20$ frames for the RDA signal. All the $6$ sequences are considered, and we observe a monotonic increasing behavior of the accuracy. Although this may not always be the case: if the initial guess of the classifier is wrong, even in the absence of tracking errors, a large value for the window would lead to a wrong classification for many frames. For this reason, a good selection approach would be to pick the lowest possible $W_h$ that guarantees a given, application dependent, accuracy target. 
For the results in \tab{tab:results}, we picked $W_h = 9$~frames, leading to a delay of $0.6$~s, as this is the lowest value of $W_h$ for which the accuracy is above $95$\% for all the sequences. Still, all values up to $W_h = 15$~frames would be good choices, as the delay is below $1$~s for all of them. The same value of $W_h$ has led to the best results also in the RD case.

	\begin{table}[t!] 
	\begin{center}
		\begin{tabular}{lccc}
			\toprule
			
			& Seq. 1 & Seq. 2 &{\bf Average}\\
			\midrule
			Acc. \% & 100.00 & 92.00 & 96.00\\
			$r_{\rm und}$ \%& 20.40 & 16.80 &18.60 \\
			$r_{\rm unk}$ \%& 0 & 0 & 0\\
			\bottomrule
		\end{tabular} 		
	\end{center}
	\caption{Accuracy results obtained in room B on two subjects using RDA processing, after training the classifier on data from room A only.}	\label{tab:diff-room}
\end{table}
\subsection{Validation in a different indoor environment}
To analyze the generalization capabilities of the proposed system, we evaluated its performance in a different environment, room B, described in \secref{sec:setup}, considering RDA processing. Here, we investigate whether a pretrained DCNN can generalize well to new rooms, as the training procedure would be too long and costly to be repeated for each new environment. The detection and tracking parameters for ghost target removal and the KF matrices can be environment independent (e.g., the KF parameters), can be easily obtained from side information (room dimensions), or can be estimated by taking some preliminary measurements on the empty room (denoising threshold). Specifically, the new range dependent denoising threshold goes from $-75$~dBm at minimum range to $-95$~dBm, which is expected given the smaller room size and the considerable presence of static objects. The threshold on the azimuth dimension is instead left unchanged.

In \tab{tab:diff-room}, we show the results of testing, in room B, the classifier trained in room A. The average accuracy for $2$ subjects over the $2$ considered sequences is $96$\%, which is lower than the one obtained in room A with $4$ subjects, but is deemed satisfactory given the difficult propagation conditions of the new environment. Indeed, we stress that room B contains furniture and several static objects which cause severe multi-path effect and clutter, in addition to the presence of other people who were working and who were not involved in the experiment. These harsh conditions are reflected in the high percentage of time in which a subject is undetectable, which is \mbox{$r_{\rm und} = 18.60$\%}, i.e., almost doubled with respect to room A (see \tab{tab:results}).
We conclude that the classifier is able to generalize to unseen environments even in realistic conditions: more reliable detection schemes would be further enhancing the model robustness and we leave their study as a future work.

\section{Conclusions}\label{sec:conclusions}

In this work, we have presented a system for indoor \mbox{multi-person} identification from \mbox{mm-wave} radar $\mu$-Doppler signatures. The proposed approach has been designed to work with \mbox{range-Doppler} (RD) and \mbox{range-Doppler-azimuth} (RDA) data, requiring only small modifications to deal with these two signals, and being able to trade working range and computational speed (RD) for detection and tracking accuracy (RDA). The processing steps are: removal of static reflections and random noise, a target detection phase using \mbox{density-based} clustering (DBSCAN), a tracking procedure using Kalman filtering and a final classification step exploiting deep convolutional neural networks (DCNNs). In our novel design, we have integrated the identification information with the trajectory tracking block. This has the twofold advantage of allowing for much higher identification accuracies when working with both RD and RDA signals in \mbox{multi-target} scenarios, i.e., where multiple subjects share and move within the same physical space. The proposed framework has been tested on real measurements involving single as well as multiple targets moving {\it concurrently} in an indoor space (a lacking aspect in the literature), obtaining an identification accuracy of $95.26$\% for RD, with $3$ targets, and of $98.27$\% with RDA, with $4$ targets. The framework has a maximum working range of $18$~m for RD and of $8$-$10$~m for RDA. 
A further evaluation was conducted to assess the generality of the proposed approach, by capturing additional test data in a different room, that was not used by the system at training time. Despite the new environment being more challenging, e.g., with furniture and other human presence, we obtained an accuracy of $96$\% with two subjects.

Future research avenues include: characterizing the indoor space by (automatically) mapping static objects and ghost reflections, which is expected to lead to higher accuracies, using multiple \mbox{time-synchronized} radar devices and 2D antenna arrays (elevation angle).

\appendices

%\section*{Acknowledgment}
%This work has been supported, in part, by MIUR (Italian Ministry of Education, University and Research) through the initiative "Departments of Excellence" (Law 232/2016) and by the EU MSCA ITN project MINTS ``MIllimeter-wave NeTworking and Sensing for Beyond 5G'' (grant no. 861222).

\bibliography{biblio}
\bibliographystyle{ieeetr}


\vspace{-1cm}

\begin{IEEEbiography}[{\includegraphics[width=1in,height=1.25in,clip,keepaspectratio]{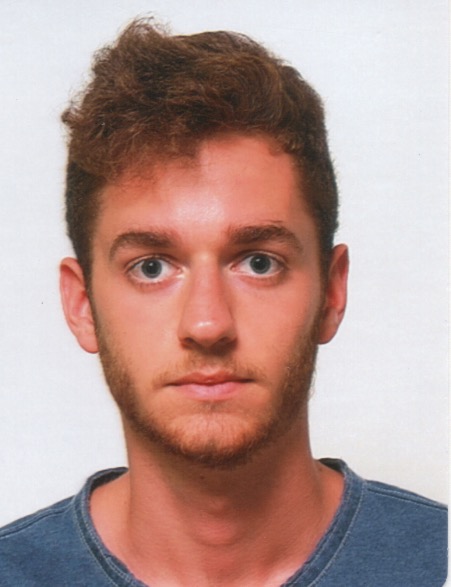}}]%
{Jacopo Pegoraro} (S'20) received the B.Sc. degree in information engineering and the M.Sc. degree in ICT for Internet and Multimedia engineering from the University of Padova, Padua, Italy, in 2017 and 2019, respectively. He is currently pursuing the Ph.D. degree with the SIGNET Research Group, Department of Information Engineering, at the same University. His research interests include deep learning and signal processing with applications to radio frequency sensing and, specifically, \mbox{mm-wave} radar sensing.
\end{IEEEbiography}

\vspace{-1cm}

\begin{IEEEbiography}[{\includegraphics[width=1in,height=1.25in,clip,keepaspectratio]{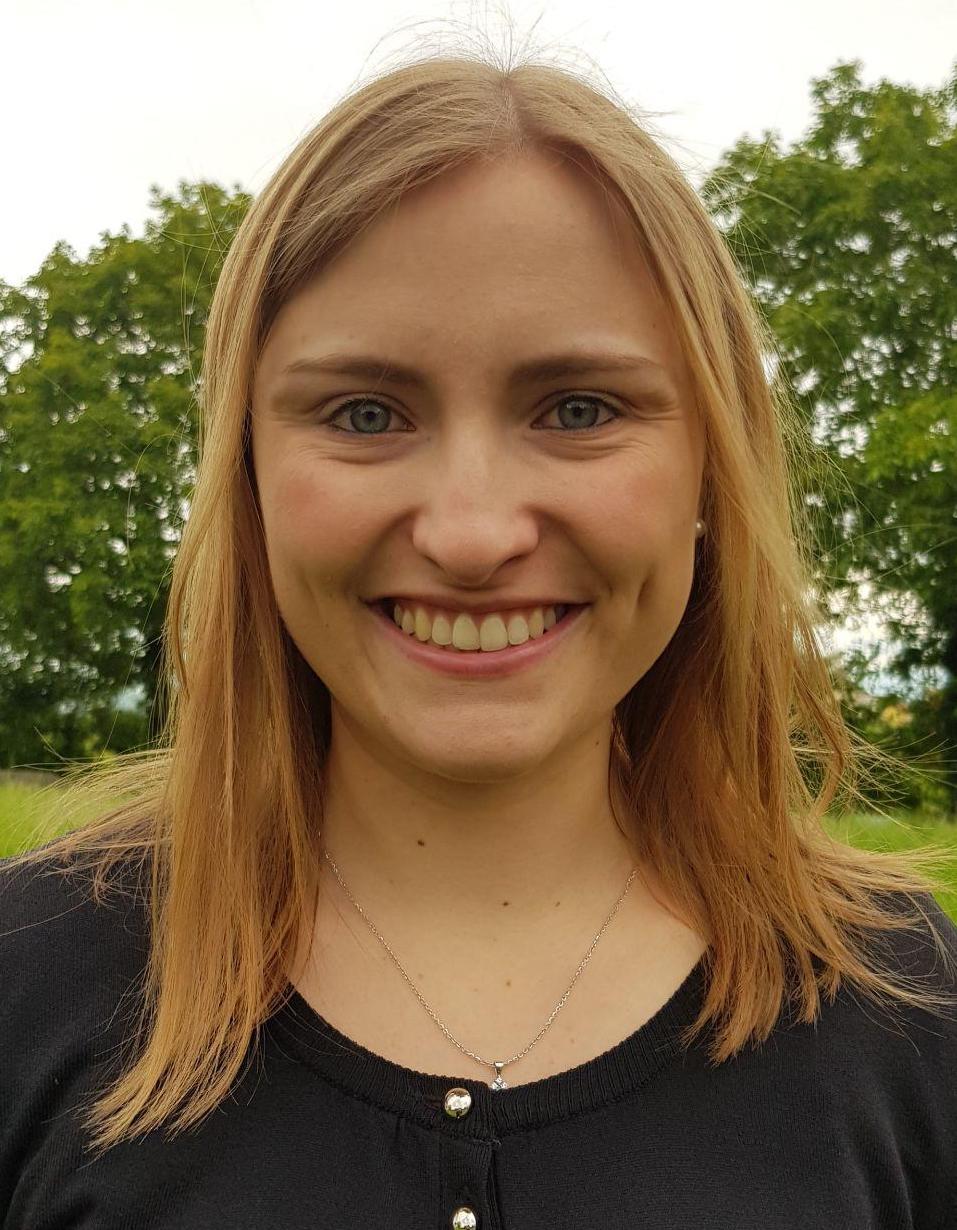}}]%
{Francesca Meneghello} (S'19) received the B.Sc. degree in information engineering and the M.Sc. degree in telecommunication engineering from the University of Padova, Italy, in 2016 and 2018 respectively. She is currently pursuing the Ph.D. degree with the Department of Information Engineering at the same university. Her current research interests include \mbox{deep-learning} architectures and signal processing with application to remote radio frequency sensing and wireless networks. She was a recipient of the Best Student Paper Award at WUWNet 2016, the Best Student Presentation Award at the IEEE Italy Section SSIE 2019 and received an honorary mention in the 2019 IEEE ComSoc Student Competition.
\end{IEEEbiography}

\vspace{-1cm}

\begin{IEEEbiography}[{\includegraphics[width=1in,height=1.25in,clip,keepaspectratio]{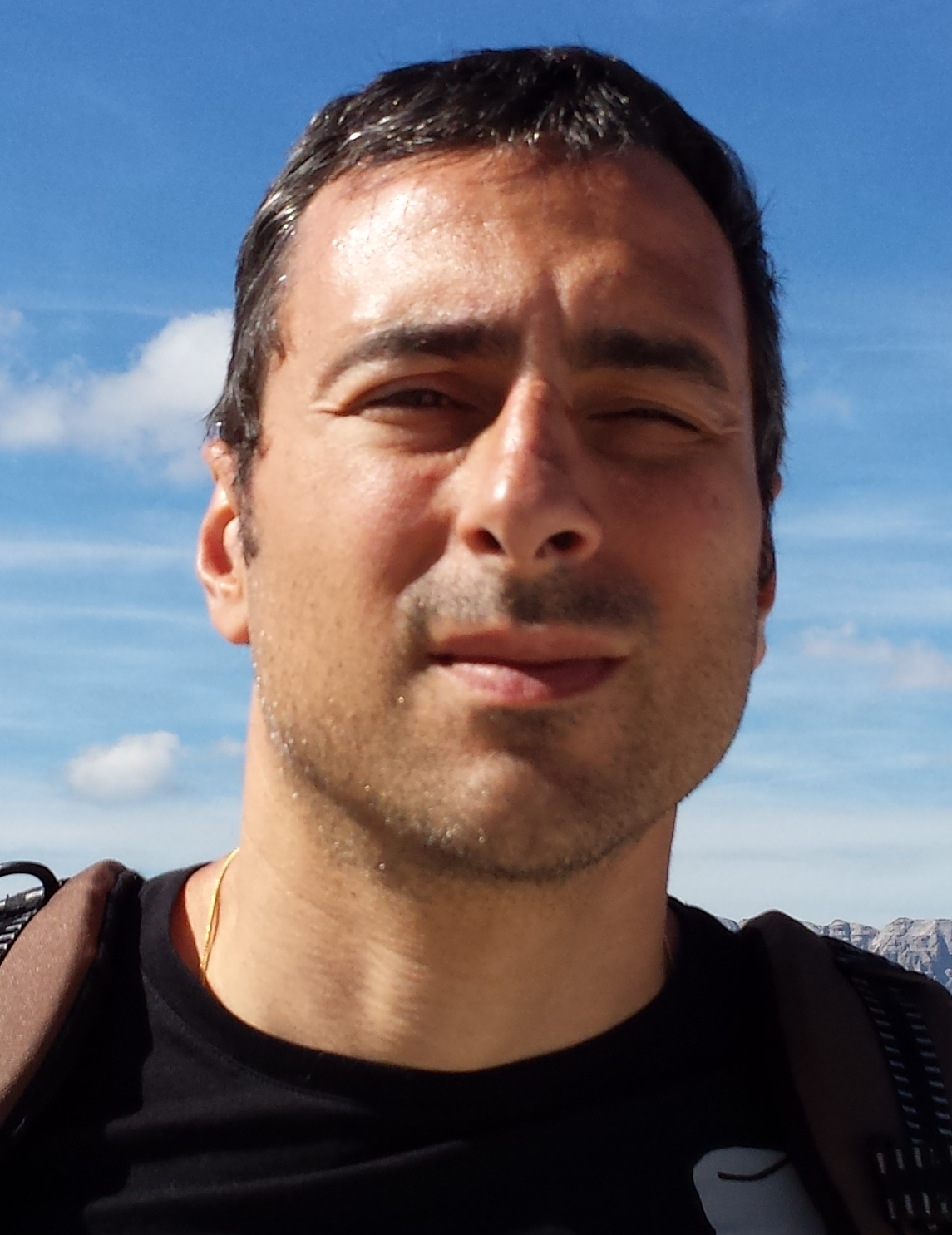}}]%
{Michele Rossi} (SM'13) is a Professor of Telecommunications in the Department of Information Engineering (DEI) at the University of Padova (UNIPD), Italy, teaching courses within the Master's Degree in ICT for internet and Multimedia (\url{http://mime.dei.unipd.it/}). He also sits on the Directive Board of the Master's Degree in Data Science offered by the Department of Mathematics (DM) at UNIPD (\url{https://datascience.math.unipd.it/}), for which he teaches machine learning and neural networks for the analysis of human data. Since 2017, he has been the Director of the DEI/IEEE Summer School of Information Engineering (\url{http://ssie.dei.unipd.it/}). His research interests lie in wireless sensing systems, green mobile networks, edge and wearable computing. In recent years, he has been involved in several EU projects on IoT technology (e.g., \mbox{IOT-A}, project no. 257521), and has collaborated with companies such as \mbox{DOCOMO} (compressive dissemination and network coding for distributed wireless networks) and Worldsensing (optimized IoT solutions for smart cities). In 2014, he has been the recipient of a SAMSUNG GRO award with a project entitled ``Boosting Efficiency in Biometric Signal Processing for Smart Wearable Devices''. In 2016-2018, he has been involved in the design of IoT protocols exploiting cognition and machine learning, as part of INTEL's Strategic Research Alliance (ISRA) R\&D program. His research is currently supported by the European Commission through the H2020 projects SCAVENGE (no. 675891) on ``green 5G networks'', MINTS (no. 861222) on ``mm-wave networking and sensing'' and GREENEDGE (no. 953775) on ``green edge computing for mobile networks'' (project coordinator). Dr. Rossi has been the recipient of seven best paper awards from the IEEE and currently serves on the Editorial Boards of the IEEE Transactions on Mobile Computing, and of the Open Journal of the Communications Society.
\end{IEEEbiography}


%\vfill


\end{document}